\documentclass[11pt]{article}
\usepackage[margin=0.9in]{geometry}
\usepackage{amsmath,amssymb,amsfonts,amsthm}
\usepackage{natbib}
\usepackage{algorithmic}
\usepackage{graphicx}
\usepackage{textcomp}
\usepackage{enumitem}  
\usepackage{graphicx}
\usepackage{xcolor}
\usepackage{float} 
\usepackage{subfigure}
\usepackage{booktabs}
\usepackage{multirow}
\usepackage{longtable}
\usepackage{array, caption, threeparttable}
\usepackage{hyperref}
\usepackage{lipsum}
\usepackage[normalem]{ulem}
\usepackage{cancel}
\usepackage{makecell}
\usepackage{lscape}
\usepackage{caption}
\allowdisplaybreaks[4]
\DeclareMathAlphabet{\mathpzc}{OT1}{pzc}{m}{it}
\usepackage{pdflscape}
\usepackage[figuresright]{rotating}
\usepackage{xr}

\usepackage{mathbbol}
\usepackage{hyperref}%

\usepackage{mathrsfs}
\usepackage[scr=stixtwofancy,cal=cm]{mathalfa} 
\DeclareMathAlphabet{\mathdutchcal}{U}{dutchcal}{m}{n}
\SetMathAlphabet{\mathdutchcal}{bold}{U}{dutchcal}{b}{n}
\DeclareMathAlphabet{\mathdutchbcal}{U}{dutchcal}{b}{n}

\theoremstyle{plain}
\newtheorem{theorem}{Theorem}

\theoremstyle{definition}

\newtheorem{remark}{Remark}

\allowdisplaybreaks[2]

\newcommand{\blind}{1}

\begin{document}

\def\spacingset#1{\renewcommand{\baselinestretch}{#1}\small\normalsize} 
\spacingset{1}


\if1\blind
{
  \title{\bf Addressing both variable selection and misclassified responses with parametric and semiparametric methods\footnotemark[2]}  
  \author{
  Hui Guo$^{1,a}$ , 
  Grace Y. Yi\footnotemark[1] $^{1, 2, b}$, 
  and 
  Boyu Wang$^{1,c}$\\
  \\
  {\footnotesize $^1$Department of Computer Science, University of Western Ontario, Ontario, Canada}\\
  {\footnotesize $^2$Department of Statistical and Actuarial Sciences, University of Western Ontario, Ontario, Canada}\\
  {\footnotesize $^a$hguo288@uwo.ca; $^b$gyi5@uwo.ca; $^c$bwang@csd.uwo.ca}
  }
  \date{} 
  \maketitle
} \fi

\renewcommand{\thefootnote}{\fnsymbol{footnote}}
\footnotetext[1]{Corresponding author.}
\footnotetext[2]{This paper has been published in \textit{Bernoulli} 32(2): 1303-1327 (May 2026). \href{https://projecteuclid.org/journals/bernoulli/volume-32/issue-2/Addressing-both-variable-selection-and-misclassified-responses-with-parametric-and/10.3150/25-BEJ1907.short}{DOI: 10.3150/25-BEJ1907.}}

\if0\blind
{
  \bigskip
  \bigskip
  \bigskip
  \begin{center}
    {\LARGE\bf Title}
\end{center}
  \medskip
} \fi

\bigskip
\begin{abstract}
While variable selection has received extensive attention in the literature, its exploration in the presence of response measurement error remains underexplored. In this paper, we investigate this important problem within the context of binary classification with error-prone responses. We present valid variable selection procedures to address the complexities of response errors. Leveraging validation data, we introduce both parametric and semiparametric methodologies to accommodate the mismeasurement effects. By rigorously establishing theoretical results, we offer insights and justifications of the validity of the proposed methods. By properly choosing {the} penalty function and regularization parameter, we demonstrate that the resulting estimators possess the oracle property. To assess the finite sample properties of the proposed methods, we conduct numerical studies that confirm the effectiveness of our proposed methods.
\end{abstract}

\noindent%
\vfill

\newpage
\spacingset{1.5} 

\section{Introduction}

In recent decades, significant progress has been made in variable selection procedures. Best subset selection is perhaps {one of} the most intuitive variable selection technique{s}, employing criteria such as the Akaike information criterion (AIC, \cite{akaike1973maximum}), the Bayesian information criterion (BIC, \cite{schwarz1978estimating}), $\phi$-criterion \cite{hannan1979determination}, and the risk inflation criterion (RIC, \cite{foster1994risk}). However, this approach becomes computationally prohibitive when the dimension of data is moderately high. {By} viewing the subset selection approach as a special case of penalized likelihood methods with $L_0$ regularization, various penalized approaches have been developed to address the computational challenges {associated with} subset selection. {Among these methods,} the most widely used techniques {include} the least absolute shrinkage and selection operator (Lasso, \cite{tibshirani1996regression}), the smoothly clipped absolute deviation (SCAD, \cite{fan2001variable}), the  minimax concave penalty (MCP, \cite{zhang2010nearly}), and their variants.

While these methods are useful {in handling variable selection, their validity relies on the critical assumption that the {{response variable}} must be accurately measured, an assumption that often does not hold for real-world applications.} Measurement error is ubiquitous in {practice \cite{carroll2006measurement, yi2017statistical, grace2021handbook}, and it is broadly differentiated as covariate measurement error and response measurement error.

Regarding variable selection with covariate measurement error,} \cite{liang2009variable} {proposed penalized} least squares and penalized quantile regression methods for partially linear models. \cite{ma2010variable} developed a framework {based on} penalized estimating equations.  \cite{loh2011high} proposed a nonconvex modification of Lasso for high-dimensional sparse linear regression with noisy, missing, and/or dependent covariates.
\cite{yi2015variable} {explored} marginal methods for simultaneous model selection and estimation for longitudinal data with missing responses and error-prone covariates. \cite{datta2017cocolasso} explored the CoCoLasso method that handles corrupted datasets. \cite{brown2019meboost} proposed the Measurement Error Boosting (MEBoost) algorithm. {\cite{chen2021analysis} proposed graphical proportional hazard models for handling survival data. \cite{yi2023estimation} developed an estimation method for the average treatment effect that simultaneously addresses variable selection and measurement error for potential confounders.}

{Although} there has been research about variable selection with measurement error in covariates, variable selection with response {measurement} error has {not been available to our best knowledge. To address this gap, in this paper, we present valid variable selection procedures that account for response measurement error, especially for the setting with misclassification in binary responses. We propose both parametric and semiparametric approaches for the scenario with a main study sample and an internal validation subsample. We introduce} both parametric and semiparametric penalized likelihood methods {and} comprehensively study the asymptotic properties of the resultant estimators. Our proposed methods can not only conduct variable selection but also address the effects induced from misclassification in responses. 

{However, these advantages come at a cost: incorporating the misclassification process introduces non-convexity into the loss function}, thus, generating a {challenging} non-convex optimization problem. To address this, {we employ the approximate regularization path-following method of \cite{wang2014optimal}, iteratively applying a variant of Nesterov’s proximal-gradient method \cite{nesterov2013gradient} to obtain approximate local solutions. A line-search strategy of \cite{bach2012optimization} is adopted to determine the learning step size. In our semiparametric approach, we use kernel-based methods for density estimation \cite{hardle1990applied, racine2004nonparametric}, while addressing the curse of dimensionality through techniques outlined in \cite{eckstein2023dimensionality}. Tuning parameters in both methods are selected using GCV or BIC, following prior works \cite{wang2007tuning, ma2010variable, fan2001variable}.} 

The remainder of this article is organized as follows. We introduce the notation and basic model setup in Section \ref{sec.2}. In Sections \ref{sec.3} and \ref{sec.3.3}, we {develop} variable selection procedures using validation
data via penalized strategies {to handle data} with misclassified responses, where we develop both parametric and semiparametric methods. {We} establish the asymptotic properties of the proposed bias correction methods. Details of implementation procedures are described in Section \ref{sec.4}, followed by simulation studies and data analysis in Section \ref{sec.5}. We conclude the article with discussions in Section \ref{sec.6}. Proofs of theoretical results, technical details, {and additional numerical results,} are provided in the Supplementary Material \cite{guo2023addressing}.

\section{Notation and model setup}\label{sec.2}

For {any} subject, let ${Y}$ denote the binary response taking value 0 or 1, and let ${Z}$ represent a $p \times 1$ random vector of associated covariates. {Let ${\mu(z)}=P(Y = 1|Z = z)$ denote the conditional mean of the response variable ${Y}$, given ${Z}=z$. Consider a regression model:
\begin{align}\label{Model}
    &{g(\mu(z))=\beta^*_0+z^{\scriptscriptstyle\mathrm{T}}\beta},
\end{align}
where $\Bar{\beta}=(\beta^*_0, \beta^{\scriptscriptstyle\mathrm{T}})^{\scriptscriptstyle\mathrm{T}}$ is the vector of unknown parameters, with $\beta\in\mathbb{R}^{p}$ representing the regression coefficients and $\beta^*_0\in\mathbb{R}$ denoting the intercept; and $g(\cdot)$ is the link function that is strictly monotone and differentiable. Common choices of $g(\cdot)$ include the {logit}, probit, and complementary log-log functions.}

We are interested in identifying active components in ${Z}$ and estimating the corresponding nonzero parameters contained in $\beta$ {using measurements of a random sample}. Let $\beta^*_{00}$ and $\beta_{0}$  represent the true {values} of $\beta^*_0$ and $\beta$, respectively, and denote $\Bar{\beta}_0=(\beta^*_{00}, \beta_{0}^{\scriptscriptstyle\mathrm{T}})^{\scriptscriptstyle\mathrm{T}}$. Without loss of generality, write $\beta_0=(\beta^{\scriptscriptstyle\mathrm{T}}_{\scriptscriptstyle\mathrm{I0}}, \beta^{\scriptscriptstyle\mathrm{T}}_{\scriptscriptstyle\mathrm{II0}})^{\scriptscriptstyle\mathrm{T}}$ so that $\beta_{\scriptscriptstyle\mathrm{I0}}$ and $\beta_{\scriptscriptstyle\mathrm{II0}}$ contain all nonzero elements and all zero components of $\beta_0$, respectively. Let $s$ denote the dimension of $\beta_{\scriptscriptstyle\mathrm{I0}}$, and we write $\Bar{\beta}_{\scriptscriptstyle\mathrm{I0}}=(\beta^*_{00}, \beta_{\scriptscriptstyle\mathrm{I0}}^{\scriptscriptstyle\mathrm{T}})^{\scriptscriptstyle\mathrm{T}}$. {Similarly, for parameter $\beta$, we express $\beta=(\beta^{\scriptscriptstyle\mathrm{T}}_{\scriptscriptstyle\mathrm{I}}, \beta^{\scriptscriptstyle\mathrm{T}}_{\scriptscriptstyle\mathrm{II}})^{\scriptscriptstyle\mathrm{T}}$ and $\Bar{\beta}_{\scriptscriptstyle\mathrm{I}}=(\beta^*_{0}, \beta_{\scriptscriptstyle\mathrm{I}}^{\scriptscriptstyle\mathrm{T}})^{\scriptscriptstyle\mathrm{T}}$, where $\beta_{\scriptscriptstyle\mathrm{I}}^{\scriptscriptstyle}$ represents the vector consisting of the first $s$ elements of $\beta$, corresponding to $\beta_{\scriptscriptstyle\mathrm{I0}}^{\scriptscriptstyle}$.}

{In the ideal situation where the variables are accurately measured, the penalized likelihood method can be applied for variable selection. However, this approach breaks down if the data are error-contaminated. In practice, the response $Y$ is often subject to misclassification. For example, in medical studies, precise diagnostic tools like biopsies can be expensive or impractical for large datasets. As a result, diagnoses are often based on less accurate methods, such as symptoms or imaging, with only a small subset of cases verified using high-precision tools. 

To address this, let $Y^*$ denote a surrogate or an observed version of $Y$. Let $\mu^*(z)=P(Y^*=1|Z=z)$ denote the conditional mean of the surrogate version $Y^*$, given covariates $Z=z$. Let
\begin{align*}
   \gamma_{10}(z) = P(Y^* = 0|Y = 1, Z=z)\ \textrm{and}\ \gamma_{01}(z) = P(Y^* = 1|Y = 0, Z=z) 
\end{align*}
represent the conditional misclassification probabilities for the response variable, given $Z=z$. As mentioned in \cite[Chapter 8]{yi2017statistical}, the conditional probabilities $\mu^*(z)$ and $\mu(z)$ are linked via
\begin{equation}
    \mu^*(z) = \gamma_{01}(z) + \{1 - \gamma_{01}(z) - \gamma_{10}(z)\}\mu(z).\label{mu_link}
\end{equation}

Consider the setting where the main study sample contains measurements $\mathcal{D}_{\text{\tiny M}}\triangleq\left\{\{Y^*_i, Z_i\}:i\in \mathcal{M}\right\}$, and a randomly selected internal validation subsample from the main study sample is also available with measurements $\mathcal{D}_{\text{\tiny V}}\triangleq\left\{\{Y_i, Y^*_i, Z_i\}:i\in \mathcal{V}\right\}$. Here, $\mathcal{M}$ and $\mathcal{V}$ stand for the subject sets for the main study and validation samples, respectively, and $\{Y_i,Y^*_i,Z_i\}$ have the same distribution as $\{Y,Y^*, Z\}$ \cite{yi2019parametric}. 
Let $n$ and $n_v$ denote the size of $\mathcal{M}$ and $\mathcal{V}$, respectively, and let $\delta=\lim_{n\rightarrow\infty}\frac{n_v}{n}$ represent the validation subsample ratio with $\delta\in(0, 1]$. This setting also features semi-supervised learning that is often discussed in machine learning research, where the focus is to train a classification procedure using the validation data $\mathcal{D}_{\text{\tiny V}}$, and then apply it to the main study data $\mathcal{D}_{\text{\tiny M}}$}.

\section{Parametric method}\label{sec.3}

{In the presence of misclassification of ${Y}$, inference about the model parameters for model (\ref{Model}) requires the attention for the misclassification probabilities {$\gamma_{01}(z)$ and $\gamma_{10}(z)$, given $Z=z$}.} We consider logistic regression models for the response misclassification {probabilities in (\ref{mu_link})}:
\begin{equation}\label{mis_pro}
    \textrm{logit}\ \gamma_{01}({z}) = g_0({z}; \nu);\ \ 
    \textrm{logit}\ \gamma_{10}({z}) = g_1({z}; \nu),
\end{equation}
where $g_0(\cdot)$ and $g_1(\cdot)$ are specified functions, and $\nu$ is a $q$-dimensional vector of unknown regression parameters. Let $\theta = (\Bar{\beta}^{\scriptscriptstyle\mathrm{T}}, \nu^{\scriptscriptstyle\mathrm{T}})^{\scriptscriptstyle\mathrm{T}}$ denote the vector of all involved parameters, and let $\theta_{0} = (\Bar{\beta}_{0}^{\scriptscriptstyle\mathrm{T}}, \nu_{0}^{\scriptscriptstyle\mathrm{T}})^{\scriptscriptstyle\mathrm{T}}$ {denote} the true value of $\theta$. Let $d=1+p+q$ denote the dimension of $\theta$. {While the logit function is used in (\ref{mis_pro}), model (\ref{mis_pro}) includes a broader class of models than logistic regression models, because functions $g_0(\cdot)$ and $g_1(\cdot)$ do not need to be restricted to be linear functions but can be of any function forms to facilitate the dependence of {misclassification} probabilities on the covariates. Furthermore, our following development is not restricted to such models only. Our methods apply to other models that allow regularity conditions detailed in Appendix {\color{blue}B1} of the Supplementary Material.} 

\subsection{Observed likelihood function}\label{sec.3.1}

With main study/validation data, likelihood function for $\theta$ can be constructed as
\begin{align*}
    L^*(\theta) = \left\{\prod_{i \in \mathcal{V}}f(y_i, y_i^*|z_i)\right\}\left\{\prod_{i \in \mathcal{M \backslash V}}f(y^*_i|z_i)\right\}, 
\end{align*}
where $f(\cdot|\cdot)$ denotes the conditional distribution of the corresponding variables with the dependence on $\theta$ suppressed in the notation. {Under the} models in Section \ref{sec.2}, following \cite[Chapter 8]{yi2017statistical}, we further write the likelihood function as 
\begin{align}
    L^*(\theta) =\prod_{i \in \mathcal{V}}&{\big[}{\mu(z_i)}^{y_i}{\left\{1-\mu(z_i)\right\}}^{1-y_i}\{{a_1(y^*_i,z_i)}\}^{y_i}\{{a_0(y^*_i,z_i)}\}^{1-y_i}{\big]} \notag\\
    &\cdot\prod_{i \in \mathcal{M \backslash  V}}{\Big[}{{\mu^*(z_i)}}^{y^*_i}{\{1-\mu^*(z_i)\}}^{1-y^*_i}{\Big]}, \notag
\end{align}
where, for $k = 0$ and $1$, ${a_k(y^*_i,z_i)} = P(Y^*_i = y^*_i|Y_i = k, {Z_i=z_i})$, given by {$a_0(y^*_i)=\gamma_{01}(z_i)^{y^*_i}\{1-\gamma_{01}(z_i)\}^{1-y^*_i}$ and $a_1(y^*_i)=\gamma_{10}(z_i)^{1-y^*_i}\{1-\gamma_{10}(z_i)\}^{y^*_i}$}.

Let 
\begin{align}\label{log_lik_para}
    \ell^*(\theta)=\log \ L^*(\theta)
\end{align}
denote the log-likelihood function for the observed data and let $S^*(\theta)=\partial \ell^*(\theta)/\partial\theta$ denote the score function. Then 
\begin{align}\label{lik_score}
    S^*(\theta)=\sum_{i \in \mathcal{V}}S_i(\theta;y_i, y^*_i, z_i) + \sum_{i \in \mathcal{M \backslash V}}S^*_i(\theta;y^*_i, z_i),
\end{align}
with $S_i(\theta;y_i, y^*_i, z_i) = (S_i^{\scriptscriptstyle\mathrm{T}}(\Bar{\beta}), S_i^{\scriptscriptstyle\mathrm{T}}(\nu))^{\scriptscriptstyle\mathrm{T}}$ for $i\in\mathcal{V}$ and $S^*_i(\theta; y^*_i, z_i) = (S_i^{*\scriptscriptstyle\mathrm{T}}(\Bar{\beta}), S_i^{*\scriptscriptstyle\mathrm{T}}(\nu))^{\scriptscriptstyle\mathrm{T}}$ for $i\in\mathcal{M\backslash V}$, where 
\begin{align}
    S_i(\Bar{\beta}) &\triangleq \left[\frac{y_i-{\mu(z_i)}}{{\mu(z_i)}{\left\{1-\mu(z_i)\right\}}}\right]{\bigg\{}\frac{\partial{\mu(z_i)}}{\partial\Bar{\beta}}{\bigg\}};\label{S_beta}\\
    S_i(\nu) &\triangleq \left\{\frac{y_i}{{a_1(y^*_i,z_i)}}\right\}\left\{\frac{\partial {a_1(y^*_i,z_i)}}{\partial\nu}\right\} + \left\{\frac{1-y_i}{{a_0(y^*_i,z_i)}}\right\}\left\{\frac{\partial {a_0(y^*_i,z_i)}}{\partial\nu}\right\};\notag\\
    S_i^*(\Bar{\beta})&\triangleq\frac{\partial}{\partial\Bar{\beta}}\log f(y^*_i|z_i);\ \ S_i^*(\nu)\triangleq\frac{\partial}{\partial\nu}\log f(y^*_i|z_i).\label{S_i*_beta} 
\end{align}

Now we express $S^*_i(\theta;y^*_i, z_i)$ in terms of models (\ref{Model}) and (\ref{mis_pro}). First, we express $S_i^*(\Bar{\beta})$ as
\begin{align} \label{S*_beta}
    S_i^*(\Bar{\beta}) &=\frac{1}{f(y^*_i|z_i)}\cdot\frac{\partial}{\partial\Bar{\beta}}f(y^*_i|z_i)\notag\\
    &=\frac{1}{\sum_{y_i=0,1}f(y^*_i|y_i,z_i)f(y_i|z_i)}\cdot\frac{\partial}{\partial\Bar{\beta}}{\bigg\{}\sum_{y_i=0,1}f(y^*_i|y_i,z_i)f(y_i|z_i){\bigg\}}\notag\\
    &=\frac{1}{{a_0(y^*_i,z_i)}{\left\{1-\mu(z_i)\right\}}+{a_1(y^*_i,z_i)}{\mu(z_i)}}\cdot\frac{\partial}{\partial\Bar{\beta}}{\big[}{a_0(y^*_i,z_i)}{\left\{1-\mu(z_i)\right\}}+{a_1(y^*_i,z_i)}{\mu(z_i)}{\big]}\notag\\
    &=\frac{{a_1(y^*_i,z_i)}-{a_0(y^*_i,z_i)}}{{a_1(y^*_i,z_i)}{\mu(z_i)}+{a_0(y^*_i,z_i)}{\left\{1-\mu(z_i)\right\}}}\frac{\partial{\mu(z_i)}}{\partial\Bar{\beta}},
\end{align}
where the third equality follows from the definitions of ${a_0(y_i^*,z_i)}$, ${a_1(y^*_i,z_i)}$ and ${\mu(z_i)}$ given in Section \ref{sec.2}, and the fourth equality holds since ${a_0(y_i^*,z_i)}$ and ${a_1(y_i^*,z_i)}$ are functionally independent of $\Bar{\beta}$. 

Similarly, for {$S_i^*(\nu)$}, we have that
\begin{align} \label{S*_nu}
    S_i^*(\nu) &=\frac{1}{{a_0(y^*_i,z_i)}{\left\{1-\mu(z_i)\right\}}+{a_1(y^*_i,z_i)}{\mu(z_i)}}\cdot\frac{\partial}{\partial\nu}\left\{{a_0(y^*_i,z_i)}{\left\{1-\mu(z_i)\right\}}+{a_1(y^*_i,z_i)}{\mu(z_i)}\right\}\notag\\
    &=\frac{1}{{a_1(y^*_i,z_i)}{\mu(z_i)}+{a_0(y^*_i,z_i)}{\left\{1-\mu(z_i)\right\}}}{\bigg[}{\left\{1-\mu(z_i)\right\}}\frac{\partial {a_0(y^*_i,z_i)}}{\partial\nu}+{\mu(z_i)}\frac{\partial {a_1(y^*_i,z_i)}}{\partial\nu}{\bigg]}\notag\\
    &=\frac{(-1)^{y^*_i+1}}{{a_1(y^*_i,z_i)}{\mu(z_i)}+{a_0(y^*_i,z_i)}{\left\{1-\mu(z_i)\right\}}}\left[\frac{\partial\gamma_{01}({z_i})}{\partial\nu}-{\mu(z_i)}\left\{\frac{\partial\gamma_{01}({z_i})}{\partial\nu}+\frac{\partial\gamma_{10}({z_i})}{\partial\nu}\right\}\right],
\end{align}
where the second equality holds since ${\mu(z_i)}$ is functionally independent of $\nu$, and the third equality comes from substituting the values 0 and 1 of $y^*_i$ respectively into the equation and utilizing the definitions of ${a_0(y_i^*,z_i)}$ and ${a_1(y^*_i,z_i)}$.

\subsection{Penalized likelihood estimator}\label{sec.3.2}

To conduct variable selection and parameter estimation simultaneously, we maximize the penalized likelihood function with respect to $\theta$:
\begin{equation}\label{Q*_ml}
    Q^*_{\scriptscriptstyle\mathrm{ML}}(\theta) = \ell^*(\theta) - n\sum_{j=1}^{p}{\rho_{\lambda}}(|\beta_j|),
\end{equation}
where $\beta_j$ is the $j$th element of $\beta$ and ${\rho_{\lambda}}(\cdot)$ is a penalty function with $\lambda$ representing a nonnegative tuning parameter.  Under regularity conditions, the maximum penalized likelihood estimator, denoted $\widehat{\theta}_{\scriptscriptstyle\mathrm{ML}}$, can be obtained by solving the penalized likelihood score equations
\begin{equation}\label{lik_eq}
    S^*(\theta)-n{\dot{\rho}_{\lambda}}(\theta)=0,
\end{equation}
where ${\dot{\rho}_{\lambda}}(\theta)=(0, {\rho'_\lambda}(|\beta_1|)\textrm{sgn}(\beta_1), ..., {\rho'_\lambda}(|\beta_p|)\textrm{sgn}(\beta_p), 0_{{q}}^{\scriptscriptstyle\mathrm{T}})^{\scriptscriptstyle\mathrm{T}}$, with ${\rho'_\lambda}(\cdot)$ {denoting} the first derivative of the penalty function ${\rho_{\lambda}}(\cdot)$, {and $\textrm{sgn}(\xi)$ is the sign function, {taking $1$ if $\xi> 0$, $0$ if $\xi=0$, and $-1$ if $\xi<0$}}. Sometimes, we loosely use $0$ to represent a scalar {(e.g., the one in ${\dot{\rho}_{\lambda}}(\theta)$)}, a vector {(e.g., the one in (\ref{lik_eq}))}, or a matrix if no confusion arises, though we {sometimes} also use $0_r$ and $0_{r_1\times r_2}$ to emphasize the dimension $r$ for a zero vector and ${r_1\times r_2}$ for a zero matrix, respectively. Here, for ease of exposition, we take the same penalty function and regularization parameter $\lambda$ for different components $\beta_j$ of $\beta$, bearing in mind that extensions for more general cases can be done. 

We now establish the asymptotic properties of the estimator $\widehat{\theta}_{\scriptscriptstyle\mathrm{ML}}$. We spell out the components of $\theta_0$ by writing $\theta_0 = ({\beta^*_{00},} \beta^{\scriptscriptstyle\mathrm{T}}_0, \nu^{\scriptscriptstyle\mathrm{T}}_0)^{\scriptscriptstyle\mathrm{T}}$ with $\beta_0=(\beta_{10},.., \beta_{{p}0})^{\scriptscriptstyle\mathrm{T}}$ and $\nu_0=(\nu_{10},..., \nu_{q0})^{\scriptscriptstyle\mathrm{T}}$. Let 
\begin{equation*}
    a_n = \textrm{max}\{|{\rho'_{\lambda_n}}(|\beta_{j0}|)|: \beta_{j0}\ne 0, \ j=1,...,{p}\}\ \text{and}\ b_n = \textrm{max}\{|{\rho''_{\lambda_n}}(|\beta_{j0}|)|: \beta_{j0}\ne 0, \ j=1,...,{p}\},
\end{equation*}
where we write $\lambda$ as $\lambda_n$ to indicate its dependence on the sample size $n$.

\begin{theorem}\label{thm1}
{Suppose that the penalty function $\rho(\cdot)$ satisfies Condition {\color{blue}B4-3} in {Appendix} {\color{blue}B}. Assume that the distributions of $\mathcal{D}_{\text{\tiny M}}$ and $\mathcal{D}_{\text{\tiny V}}$ satisfy the regularity conditions outlined in {\color{blue}B1} of the Supplementary Material.} If $b_n\rightarrow 0$ and $\lambda_n\rightarrow 0$ as $n\rightarrow 0$, then there exists a local maximizer $\widehat{\theta}_{\scriptscriptstyle\mathrm{ML}}$ of $Q^*_{\scriptscriptstyle\mathrm{ML}}(\theta)$ such that 
\begin{align*}
    \|\widehat{\theta}_{\scriptscriptstyle\mathrm{ML}}-\theta_0\|_2=O_p(n^{-1/2}+a_n),
\end{align*}
where {$\|a\|_2\triangleq \sqrt{\sum_{j=1}^{r}a^2_j}$} represents the $L_2$ norm for vector $a\triangleq(a_1,...,a_r)^{\scriptscriptstyle\mathrm{T}}$.
\end{theorem}

Theorem \ref{thm1} implies that a $\sqrt{n}$-consistent estimator of $\theta$ exists if $a_n=O(n^{-1/2})$. Next we establish the sparsity property and the asymptotic normality of the penalized likelihood estimator $\widehat{\theta}_{\scriptscriptstyle\mathrm{ML}}$. Define
\begin{align*}
    &b=(0, {\rho'_{\lambda_n}}(|\beta_{10}|)\textrm{sgn}(\beta_{10}),...,{\rho'_{\lambda_n}}(|\beta_{{ s}0}|)\textrm{sgn}(\beta_{{ s}0}))^\textrm{{\tiny T}};\ \ \Tilde{b}=(b^\textrm{{\tiny T}},0_{{q}}^\textrm{{\tiny T}})^\textrm{{\tiny T}};\\
    &\Sigma=\textrm{diag}\{0, {\rho''_{\lambda_n}}(|\beta_{10}|),...,{\rho''_{\lambda_n}}(|\beta_{{ s}0}|)\};\ \ \Tilde{\Sigma}=\begin{pmatrix}\Sigma & 0_{({ s}+1)\times {q}}\\0_{{q}\times ({ s}+1)} & 0_{{q}\times{q}}\end{pmatrix},
\end{align*}
{where $0$ in $b$ is a scalar.}

Let 
\begin{align*}
    &I^*_{1}(\theta_0)=E\left[\left\{\frac{\partial}{\partial \theta}\textrm{log}f(Y,Y^*|Z)\right\}\left\{\frac{\partial}{\partial \theta}\textrm{log}f(Y,Y^*|Z)\right\}^{\scriptscriptstyle\mathrm{T}}\right]\bigg|_{\theta = \theta_0}\triangleq\ \ \begin{pmatrix}I_{1\Bar{\beta}\Bar{\beta}}(\theta_0) & I_{1\Bar{\beta}\nu}(\theta_0)\\I_{1\nu\Bar{\beta}}(\theta_0) & I_{1\nu\nu}(\theta_0)\end{pmatrix};\\
    &I^*_{2}(\theta_0)=E\left[\left\{\frac{\partial}{\partial \theta}\textrm{log}f(Y^*|Z)\right\}\left\{\frac{\partial}{\partial \theta}\textrm{log}f(Y^*|Z)\right\}^{\scriptscriptstyle\mathrm{T}}\right]\bigg|_{\theta = \theta_0}\triangleq\ \ \begin{pmatrix}I_{2\Bar{\beta}\Bar{\beta}}(\theta_0) & I_{2\Bar{\beta}\nu}(\theta_0)\\I_{2\nu\Bar{\beta}}(\theta_0) & I_{2\nu\nu}(\theta_0)\end{pmatrix},
\end{align*}
where $I_{1\Bar{\beta}\Bar{\beta}}(\theta_0)$ and $I_{2\Bar{\beta}\Bar{\beta}}(\theta_0)$ stand for the $({p}+1)\times ({p}+1)$ submatrices of $I^*_{1}(\theta_0)$ and $I^*_{2}(\theta_0)$ corresponding to $\Bar{\beta}$, respectively, $I^*_{1\nu\Bar{\beta}}(\theta_{0})=I^*_{1\Bar{\beta}\nu}(\theta_{0})^{\textrm{\tiny T}}$, and $I^*_{2\nu\Bar{\beta}}(\theta_{0})=I^*_{2\Bar{\beta}\nu}(\theta_{0})^{\textrm{\tiny T}}$. Their explicit expressions under models (\ref{Model}) and (\ref{mis_pro}) are given in Appendix {\color{blue}D1} of the Supplementary Material. 

Denote $\varpi=(\Bar{\beta}_{{\scriptscriptstyle\mathrm{I}}}^{\scriptscriptstyle\mathrm{T}},\nu^{\scriptscriptstyle\mathrm{T}})^{\scriptscriptstyle\mathrm{T}}$ and $\varpi_0=(\Bar{\beta}_{{\scriptscriptstyle\mathrm{I0}}}^{\scriptscriptstyle\mathrm{T}},\nu_{\scriptscriptstyle\mathrm{0}}^{\scriptscriptstyle\mathrm{T}})^{\scriptscriptstyle\mathrm{T}}$. Let $I^*_1(\varpi_0)$ and $I^*_2(\varpi_0)$ respectively represent the submatrices of $I^*_{1}(\theta_{0})$ and $I^*_{2}(\theta_{0})$ corresponding to $\varpi_0$, and define
\begin{align}\label{I_delta}
    I^*_{\delta}(\varpi_{0})=\delta I^*_1(\varpi_{0})+(1-\delta)I^*_2(\varpi_{0}).
\end{align}
Denote $I^*_{\delta}(\varpi_{0})=\left(\begin{smallmatrix}I_{11}&I_{12}\\I_{21}&I_{22}\end{smallmatrix}\right)$ with $I_{11}$ standing for the $({ s}+1)\times ({ s}+1)$ submatrix of $I^*_{\delta}(\varpi_{0})$ corresponding to $\Bar{\beta}_{\mathrm{\scriptscriptstyle I0}}$, and define $I^*_{11}=I_{11}-I_{12}I_{22}^{-1}I_{21}$.

\begin{theorem}[Oracle Property]\label{thm2}
{Suppose the conditions in Theorem \ref{thm1} hold. Assume that $a_n=O(n^{-1/2})$.} Let $\widehat{\theta}_{\scriptscriptstyle\mathrm{ML}}\triangleq(\widehat{\Bar{\beta}}_{\scriptscriptstyle\mathrm{ML\_I}}^{\scriptscriptstyle\mathrm{T}}, \widehat{\beta}^{\scriptscriptstyle\mathrm{T}}_{\scriptscriptstyle\mathrm{ML\_II}}, {\widehat{\nu}}^{\scriptscriptstyle\mathrm{T}})^{\scriptscriptstyle\mathrm{T}}$ denote a $\sqrt{n}$-consistent estimator of $\theta$ obtained from solving equation (\ref{lik_eq}), {with $\widehat{\Bar{\beta}}_{\scriptscriptstyle\mathrm{ML\_I}}^{\scriptscriptstyle\mathrm{T}}$ and $ \widehat{\beta}^{\scriptscriptstyle\mathrm{T}}_{\scriptscriptstyle\mathrm{ML\_II}}$
estimating $\Bar{\beta}_{\scriptscriptstyle\mathrm{I0}}$ and $\beta_{\scriptscriptstyle\mathrm{II0}}$, respectively,} and let $\widehat{\varpi}=(\widehat{\Bar{\beta}}_{{\scriptscriptstyle\mathrm{ML\_I}}}^{\scriptscriptstyle\mathrm{T}},\widehat{\nu}^{\scriptscriptstyle\mathrm{T}})^{\scriptscriptstyle\mathrm{T}}$. If 
\begin{equation}
    \liminf\limits_{n\rightarrow\infty}\liminf\limits_{\xi\rightarrow0^+}\sqrt{n}{\rho'_{\lambda_n}}(\xi)=\infty,\notag
\end{equation}
then {as $n\rightarrow\infty$,}
\begin{enumerate}[label=(\roman*), leftmargin=15pt]
    \item (Sparsity) {with probability tending to 1,} $\widehat{\beta}_{\scriptscriptstyle\mathrm{ML\_II}}=0$,
    \item (Asymptotic normality) as $n\rightarrow\infty$,
    \begin{align*}
        \sqrt{n}\left\{(\widehat{\varpi}-\varpi_{0})+\{I^*_{\delta}(\varpi_{0})+\Tilde{\Sigma}\}^{-1}\tilde{b}\right\}\stackrel{d}{\longrightarrow}\mathcal{N}\left\{0,\{I^*_{\delta}(\varpi_{0})+\Tilde{\Sigma}\}^{-1}I^*_{\delta}(\varpi_{0})\{I^*_{\delta}(\varpi_{0})+\Tilde{\Sigma}\}^{-1}\right\},
    \end{align*}
    and further, 
    \begin{align*}
        \sqrt{n}\left\{(\widehat{\Bar{\beta}}_{\scriptscriptstyle\mathrm{ML\_I}}-\Bar{\beta}_{\scriptscriptstyle\mathrm{I0}})+(I^*_{11}+\Sigma)^{-1}b\right\}
        \stackrel{d}{\longrightarrow}\mathcal{N}\left\{0,(I^*_{11}+\Sigma)^{-1}I^*_{11}(I^*_{11}+\Sigma)^{-1}\right\}.
    \end{align*}
\end{enumerate}
\end{theorem}

\begin{remark}\label{remark3}
For the Lasso penalty \cite{tibshirani1996regression} with ${\rho_{\lambda_n}}(|{\xi}|)=\lambda_n |{\xi}|$, we have that $a_n=\lambda_n$ and $b_n=0$. Theorem \ref{thm1} says that $\lambda_n=O(n^{-1/2})$ is required to ensure the resulting estimator to be $\sqrt{n}$-consistent. However, such a requirement contradicts the condition $\sqrt{n}\lambda_n\rightarrow\infty$ as $n\rightarrow \infty$ in Theorem \ref{thm2}. Thus, the oracle property does generally not hold if the Lasso penalty is chosen.

{On the contrary, for} the SCAD penalty \cite{fan2001variable}, defined by ${\rho}'(|\xi|)=\lambda_n\big\{I(|\xi|\le \lambda_n)+\frac{(a\lambda_n-|\xi|)_{+}}{(a-1)\lambda_n}I(|\xi|>\lambda_n)\big\}$ for some $a>2$, we have that $a_n=0$ and $b_n=0$ as $\lambda_n\rightarrow0$, where $I(\cdot)$ is the indicator function. Hence, the resulting estimator has both $\sqrt{n}$-consistency and the oracle property if $\sqrt{n}\lambda_n\rightarrow\infty$ as $n\rightarrow\infty$. Further, if $\lambda_n$ is small enough, we have that $b=0$ and $\Sigma=0$, indicating that the asymptotic covariance matrix of $\widehat{\varpi}$ can be well approximated by $\frac{1}{n}I^*_{\delta}(\varpi_{0})^{-1}$, which is the asymptotic covariance matrix of the maximum likelihood estimator of $\varpi_{0}$ for the case with $\beta_{\scriptscriptstyle\mathrm{II0}}$ known to be zero. This approximation suggests that the estimator $\widehat{\theta}_{\scriptscriptstyle\mathrm{ML}}$ performs nearly as well as if $\beta_{\scriptscriptstyle\mathrm{II0}}=0$ were known in advance. {Similarly, for the MCP \cite{zhang2010nearly} defined as $\rho'(|\xi|)=\left(\lambda_n-\frac{|\xi|}{a}\right)I(|\xi|\le a\lambda_n)$ for some $a>0$,  {we also have $a_n=0$ and $b_n=0$ as $\lambda_n\rightarrow0$. The regularity conditions in Theorems \ref{thm1}-\ref{thm2} on the penalty function are satisfied if} $\lambda_n\rightarrow 0$ and $\sqrt{n}\lambda_n\rightarrow\infty$ as $n\rightarrow\infty$.}
\end{remark}

\begin{remark}
{
In formulating the penalized likelihood function (\ref{Q*_ml}), we focus on penalizing the regression parameter with nuisance parameter $\nu$ left unpenalized. If needed, we can also incorporate the sparsity of the nuisance parameter $\nu$ by modifying  (\ref{Q*_ml}) as $Q^{*\text{\textdagger}}_{\scriptscriptstyle\mathrm{ML}}(\theta) = \ell^*(\theta) - n\Big\{\sum_{j=1}^{p}{\rho_{\lambda}}(|\beta_j|)+$ $\sum_{j=1}^{q}{\rho_{\lambda}}(|\nu_j|)\Big\}$. Under this formulation, Theorems \ref{thm1} and \ref{thm2} remain valid, with relevant quantities modified accordingly.} 
\end{remark}

\begin{remark}
While the use of penalty functions may often shrink penalized parameters towards zero, not all penalty functions are capable of performing variable selection, as discussed by {\cite{fan2001variable} and \cite{tibshirani1996regression}}. To achieve sparsity, we require ${\rho_{\lambda}}(\cdot)$ to satisfy certain regularity conditions (as specified in Theorems~\ref{thm2} and \ref{thm5}) that ensure the estimated subvector of $\beta$ corresponding to noninformative covariates to be exactly zero. These conditions can be satisfied by commonly used penalty functions, such as SCAD and MCP.
\end{remark}

\section{Semiparametric method}\label{sec.3.3}

The penalized likelihood method in Section \ref{sec.3.2} is readily implemented. However, it requires modeling the misclassification process parametrically as in (\ref{mis_pro}). When correctly specifying the form of $\gamma_{01}({z_i})$ and $\gamma_{10}({z_i})$ is in doubt, the resulting estimator may be biased. To circumvent this issue, here we develop a semiparametric method which allows us not to place any specific structure for the misclassification process and utilize the validation sample to estimate $\gamma_{01}({z_i})$ and $\gamma_{10}({z_i})$ nonparametrically. Without loss of generality, we assume that $\gamma_{01}({z_i})$ and $\gamma_{10}({z_i})$ are functionally independent of $\Bar{\beta}$.

For generality, we write $Z$ by separating its continuous and discrete components, respectively denoted by $Z^{\textrm{\tiny C}}$ and $Z^{\textrm{\tiny D}}$, i.e., $Z=(Z^{\textrm{\tiny CT}}, Z^{\textrm{\tiny DT}})^{\scriptscriptstyle\mathrm{T}}$, where $Z^{\textrm{\tiny C}}$ and $Z^{\textrm{\tiny D}}$ can be null. Similarly, for {$z\in\mathbb{R}^{p}$} and observation $z_i$, we write $z=(z^{\textrm{\tiny CT}}, z^{\textrm{\tiny DT}})^{\scriptscriptstyle\mathrm{T}}$ and $z_i=(z_i^{\textrm{\tiny CT}}, z_i^{\textrm{\tiny DT}})^{\scriptscriptstyle\mathrm{T}}$. Let ${p_1}$ and ${p_2}$ denote the dimensions of $Z^{\textrm{\tiny C}}$ and $Z^{\textrm{\tiny D}}$, respectively, which are nonnegative integers {satisfying ${p_1}+{p_2}=p$}. Here we employ the estimation method of \cite{racine2004nonparametric} who considered mixed variables with continuous and discrete components.

To be specific, for the continuous variables, define 
\begin{align}\label{K_b}
    K_h(z_i^{\textrm{\tiny C}},z^{\textrm{\tiny C}})=h^{-{p_1}}K\{(z_i^{\textrm{\tiny C}}-z^{\textrm{\tiny C}})/h\},
\end{align}
where $K(\cdot): \mathbb{R}^{{p_1}}\rightarrow \mathbb{R}$ is a ${p_1}$-dimensional symmetric product kernel function, and $h$ is the bandwidth that can be dependent on the sample size $n$. For the discrete variables, define
\begin{align}\label{l_omega}
    l_{\omega}(z_{it}^{\textrm{\tiny D}}, z_{t}^{\textrm{\tiny D}})=
    \left\{
    \begin{aligned}
    &1, \ \ \ \ \  \textrm{if}\ z_{it}^{\textrm{\tiny D}}=z_{t}^{\textrm{\tiny D}}, \\
    &\omega, \ \ \ \ \ \textrm{if}\ z_{it}^{\textrm{\tiny D}}\ne z_{t}^{\textrm{\tiny D}},
    \end{aligned}
    \right.
\end{align}
where $z_{it}^{\textrm{\tiny D}}$ and $z_{t}^{\textrm{\tiny D}}$ denote the $t$th component of $z_i^{\textrm{\tiny D}}$ and $z^{\textrm{\tiny D}}$ for $t=1, ..., {p_2}$, respectively, and $\omega$ is a smoothing parameter taking values in $[0, 1]$. When $\omega=0$, function $l_{\omega}(z_{it}^{\textrm{\tiny D}}, z_{t}^{\textrm{\tiny D}})$ becomes an indicator function. Then the product kernel for the discrete variables is defined by 
\begin{align*}
    L_{\omega}(z_i^{\textrm{\tiny D}}, z^{\textrm{\tiny D}})=\prod_{t=1}^{{p_2}} l_{\omega}(z_{it}^{\textrm{\tiny D}}, z_{t}^{\textrm{\tiny D}})=\omega^{d_{i,z}},
\end{align*}
where $d_{i,z}=\sum_{t=1}^{{p_2}}I(z_{it}^{\textrm{\tiny D}}\ne z_{t}^{\textrm{\tiny D}})$ with $I(\cdot)$ representing the indicator function. Define
\begin{align}\label{tilde_K}
    \tilde{K}_{i,z}=\big\{K_h(z_i^{\textrm{\tiny C}},z^{\textrm{\tiny C}})\big\}^{I({p_1}\ne 0)}\big\{L_{\omega}(z_i^{\textrm{\tiny D}}, z^{\textrm{\tiny D}})\big\}^{I({p_2}\ne 0)}.
\end{align}
Then using the measurements in the validation sample $\mathcal{V}$, we estimate the misclassification probabilities by 
\begin{align}\label{ga_10_01}
    \widehat{\gamma}_{10}(z)=\frac{\sum_{i\in \mathcal{V}}y_i(1-y^*_i)\tilde{K}_{i,z}}{\sum_{i\in \mathcal{V}}y_i\tilde{K}_{i,z}};\ \ \widehat{\gamma}_{01}(z)=\frac{\sum_{i\in \mathcal{V}}(1-y_i)y^*_i\tilde{K}_{i,z}}{\sum_{i\in \mathcal{V}}(1-y_i)\tilde{K}_{i,z}}.
\end{align}
The estimates in (\ref{ga_10_01}) require the specification of $\omega$ {in (\ref{l_omega})} and $h$ and $K(\cdot)$ in (\ref{K_b}). The smoothing parameter $\omega$ may be chosen together with $h$ by cross validation or based on using statistics described in Section \ref{sec.4}. Regarding the specification of $K(\cdot)$, the second order kernel function, such as the Gaussian kernel, is often considered, where the density function of $Z^{\textrm{\tiny C}}$ is commonly assumed to be twice continuously differentiable and $K(\cdot)$ is a smooth symmetric kernel function with a compact support satisfying $\int K(t)t^2 dt=1$. Generally, if the density of $Z^{\textrm{\tiny C}}$ is $m$ times continuously differentiable with $m\ge 2$, then $K(\cdot)$ may be taken as the $m$th order kernel \cite[Chapter 24]{van2000asymptotic} {satisfying}
\begin{align*}
    \int K(u)du=1;\ \ \int u^jK(u)du=0\ \textrm{for}\ j=1,..,m-1; \int |u|^m K(u)du<\infty;\ \ \int K^2(u)du<\infty.
\end{align*}
Accordingly, ${\mu^*(z_i)}$ in (\ref{mu_link}) and ${a_k(y^*_i,z_i)}$ for $k=0, 1$ can be estimated using (\ref{ga_10_01}), denoted as ${\widehat{\mu}^*(z_i)}$ and ${\widehat{a}_k(y^*_i,z_i)}$, respectively.

Next, let $\widehat{\ell}^*(\Bar{\beta})$ denote the modified {version of the} log-likelihood function $\ell^*(\theta)$ defined in (\ref{log_lik_para}), with ${\mu^*(z_i)}$ and ${a_k(y^*_i,z_i)}$ replaced by ${\widehat{\mu}^*(z_i)}$ and ${\widehat{a}^*_k(y^*_i,z_i)}$, respectively, for $k=0, 1$,  where $\Bar{\beta}$ is the only parameter involved in the resulting function $\widehat{\ell}^*(\Bar{\beta})$. That is,
\begin{align}\label{semi_log_lik}
    \widehat{\ell}^*(\Bar{\beta}) =& \sum_{i \in \mathcal{V}}{\big[}y_i\textrm{log}\ {\mu(z_i)}+(1-y_i)\textrm{log}{\left\{1-\mu(z_i)\right\}}+y_i\textrm{log}\{{\widehat{a}_1(y^*_i,z_i)}\}+(1-y_i)\textrm{log}\{{\widehat{a}_0(y^*_i,z_i)}\}{\big]}\notag\\
    &+\sum_{i \in \mathcal{M \backslash V}}{\big[}y^*_i\textrm{log}\ {\widehat{\mu}^*(z_i)} +(1-y^*_i)\textrm{log}{\{1-\widehat{\mu}^*(z_i)\}}{\big]}.
\end{align}

Consequently, define the penalized semiparametric likelihood function with regard to $\Bar{\beta}${:}
\begin{align*}
    Q^*_{\scriptscriptstyle\mathrm{sp}}(\Bar{\beta})=\widehat{\ell}^*(\Bar{\beta})-n\sum_{j=1}^{p}{\rho_{\lambda}}(|\beta_j|),
\end{align*}
and we obtain the penalized semiparametric estimator $\widehat{\Bar{\beta}}_{\scriptscriptstyle\mathrm{sp}}$ by maximizing $Q^*_{\scriptscriptstyle\mathrm{sp}}(\Bar{\beta})$ with respect to $\Bar{\beta}$:
\begin{align}\label{beta_sp_es}
    \widehat{\Bar{\beta}}_{\scriptscriptstyle\mathrm{sp}}=\arg\max_{\Bar{\beta}}\ Q^*_{\scriptscriptstyle\mathrm{sp}}(\Bar{\beta}).
\end{align}
Analogous to (\ref{lik_eq}), under regularity conditions, $\widehat{\Bar{\beta}}_{\scriptscriptstyle\mathrm{sp}}$ can be found by solving 
\begin{equation}
    \widehat{S}^*(\Bar{\beta})-n{\dot{\rho}_{\lambda}}(\Bar{\beta})=0\label{semi_eq}
\end{equation}
for $\Bar{\beta}$, where ${\dot{\rho}_{\lambda}}(\Bar{\beta})=(0, {\rho'_\lambda}(|\beta_1|)\textrm{sgn}(\beta_1), ..., {\rho'_\lambda}(|\beta_{p}|)\textrm{sgn}(\beta_{p}))^{\scriptscriptstyle\mathrm{T}}$ and
\begin{equation}\label{S*_hat}
    \widehat{S}^*(\Bar{\beta})=\sum_{i \in \mathcal{V}}S_{i}(\Bar{\beta};y_i,y^*_i,z_i)+\sum_{i \in \mathcal{M\backslash V}}\widehat{S}^*_{i}(\Bar{\beta};y^*_i,z_i),
\end{equation}
with $S_{i}(\Bar{\beta};y_i,y^*_i,z_i)$ given by (\ref{S_beta}) and $\widehat{S}^*_{i}(\Bar{\beta};y^*_i,z_i)$ given by (\ref{S*_beta}) in which $\gamma_{01}({z_i})$ and $\gamma_{10}({z_i})$ are substituted by $\widehat{\gamma}_{01}({z_i})$ and $\widehat{\gamma}_{10}({z_i})$, respectively.

Modifying the proof of Theorem \ref{thm1}, we {can establish} the following theorem.


\begin{theorem}\label{thm4} 
Suppose that the conditions in Theorem \ref{thm1} and Conditions {\color{blue}B2-1} and {\color{blue}B2-2} in Appendix {\color{blue}B} in the Supplementary Material hold. Assume that the bandwidth $h$ and smoothing parameter $\omega$ satisfy: $h\rightarrow 0$, $nh^{2{p_1}}\rightarrow\infty$, $nh^{2m}\rightarrow0$, $\omega\rightarrow 0$, $n\omega^2\rightarrow0$, and $n_v=\delta n O(h^m+\omega)$, where $\delta$ and $n_v$ are defined in Section \ref{sec.2}. Then, there exists a root of (\ref{semi_eq}), denoted $\widehat{\Bar{\beta}}_{\scriptscriptstyle\mathrm{sp}}$, such that {as $n\rightarrow\infty$, with probability tending to 1,} 
$$\|\widehat{\Bar{\beta}}_{\scriptscriptstyle\mathrm{sp}}-\Bar{\beta}_0\|_2=O_p(n^{-1/2}+a_n).$$
\end{theorem}

The sparsity property of the estimator $\widehat{\Bar{\beta}}_{\scriptscriptstyle\mathrm{sp}}$ can be obtained in a similar way to Theorem \ref{thm2} (i). However, the asymptotic covariance matrix of $\widehat{\Bar{\beta}}_{\scriptscriptstyle\mathrm{sp}}$ in the semiparametric setting is more complicated than that of the estimator in the parametric setting. Define
\begin{align}\label{Sigma_sp}
    \Sigma_{\scriptscriptstyle\mathrm{sp}}=E\Bigg(\bigg\{\frac{\{1-\gamma_{10}(Z)-\gamma_{01}(Z)\}}{{\mu^*(Z)}{\{1-\mu^*(Z)\}}}\Big[Y_i\{1-Y^*-\gamma_{10}(Z)\}-(1-Y)\{Y^*-\gamma_{01}(Z)\}\Big]\frac{\partial {\mu(Z)}}{\partial \Bar{\beta}_{\scriptscriptstyle\mathrm{I}}}\bigg\}^{\otimes 2}\Bigg),
\end{align}
where $a^{\otimes 2}=aa^{\scriptscriptstyle\mathrm{T}}$ for any column vector $a$.

\begin{theorem}[Oracle Property]\label{thm5}
Suppose that the conditions in Theorem \ref{thm4} are satisfied and that $a_n=O(n^{-1/2})$. If
\begin{equation}
    \liminf\limits_{n\rightarrow\infty}\liminf\limits_{\xi\rightarrow0^+}\sqrt{n}{\rho'_{\lambda_n}}(\xi)=\infty,\notag
\end{equation}
then, {for} any $\sqrt{n}$-consistent estimator $\widehat{\Bar{\beta}}_{\scriptscriptstyle\mathrm{sp}}\triangleq({\widehat{\Bar{\beta}}_{\scriptscriptstyle\mathrm{sp\_I}}}^{\scriptscriptstyle\mathrm{T}}, {\widehat{\beta}_{\scriptscriptstyle\mathrm{sp\_II}}}^{\scriptscriptstyle\mathrm{T}})^{\scriptscriptstyle\mathrm{T}}$ obtained from solving equation (\ref{semi_eq}), {the following results hold: as $n\rightarrow\infty$,}
\begin{enumerate}[label=(\roman*), leftmargin=15pt]
    \item (Sparsity) {with probability tending to 1,} $\widehat{\beta}_{\scriptscriptstyle\mathrm{sp\_II}}=0$,
    \item (Asymptotic normality) 
    \begin{align*}
        \sqrt{n}\left\{I_{11}+\Sigma\right\}\left[(\widehat{\Bar{\beta}}_{\scriptscriptstyle\mathrm{sp\_I}}-\Bar{\beta}_{\scriptscriptstyle\mathrm{I0}})+\left\{I_{11}+\Sigma\right\}^{-1}b\right]{\stackrel{d}{\longrightarrow}}\mathcal{N}\left[0,I_{11}+\frac{(1-\delta)^2}{\delta}\Sigma_{\scriptscriptstyle\mathrm{sp}}\right],
    \end{align*}
    where $I_{11}$ {denotes} the $({ s}+1)\times ({ s}+1)$ submatrix of $I^*_{\delta}(\varpi_{0})$ corresponding to $\Bar{\beta}_{\mathrm{\scriptscriptstyle I0}}$ defined after (\ref{I_delta}).
\end{enumerate}
\end{theorem}

\begin{remark}\label{remark6}
If ${p_1}=0$, then vector $Z$ is composed entirely of discrete variables. In this case, $\tilde{K}_{i,z}=L_{\omega}(z_i^{\textrm{\tiny D}}, z^{\textrm{\tiny D}})$, and the restriction on the bandwidth $h$ in Condition {\color{blue}B3-2} in {Appendix} {\color{blue}B} in the Supplementary Material is not needed. On the other hand, if ${p_2}=0$, then $Z$ contains only continuous variables. (\ref{tilde_K}) becomes $\tilde{K}_{i,z}=K_h(z_i^{\textrm{\tiny C}}, z^{\textrm{\tiny C}})$, and the requirement for the smoothing parameter $\omega$ in Condition {\color{blue}B3-3} in {Appendix} {\color{blue}B} in the Supplementary Material is not needed. 

\end{remark}

For ease of exposition, let $\widehat{\Bar{\beta}}_{\scriptscriptstyle\mathrm{I}}=(\widehat{\beta}^*_{0}, \widehat{\beta}_{\scriptscriptstyle\mathrm{I}}^{\scriptscriptstyle\mathrm{T}})^{\scriptscriptstyle\mathrm{T}}$ represent an estimator of $\Bar{\beta}_{\scriptscriptstyle\mathrm{I}}$, which is either $\widehat{\Bar{\beta}}_{\scriptscriptstyle\mathrm{ML\_I}}$ under the parametric setting (\ref{mis_pro}) or $\widehat{\Bar{\beta}}_{\scriptscriptstyle\mathrm{sp\_I}}$ under the semiparametric setting (\ref{ga_10_01}). The asymptotic covariance matrix of $\widehat{\Bar{\beta}}_{\scriptscriptstyle\mathrm{I}}$ may be estimated using Theorem{} \ref{thm2} or \ref{thm5}, depending on whether (\ref{mis_pro}) or (\ref{ga_10_01}) is employed. To be specific, let $\Sigma_{\lambda}(\widehat{\beta}_{\scriptscriptstyle\mathrm{I}})=\textrm{diag}\{ {\rho}''_{\lambda}(|\widehat{\beta}_1|),...,{\rho}''_{\lambda}(|\widehat{\beta}_{ s}|)\}$ and $\Sigma_{\lambda}(\widehat{\Bar{\beta}}_{\scriptscriptstyle\mathrm{I}})=\textrm{diag}\{0, {\rho}''_{\lambda}(|\widehat{\beta}_1|),...,{\rho}''_{\lambda}(|\widehat{\beta}_{ s}|)\}$. Let $\widehat{I}_{11}$, $\widehat{I}_{12}$ and $\widehat{I}_{22}$ represent the sample approximation of $I_{11}$, $I_{12}$ and $I_{22}$ with $\theta_{0}$ replaced by $\widehat{\theta}_{\mathrm{\scriptscriptstyle ML}}$, and denote $\widehat{I}^*_{11}=\widehat{I}_{11}-\widehat{I}_{12}\widehat{I}_{22}^{{-\textrm{\scriptsize 1}}}\widehat{I}_{12}^{\textrm{\tiny T}}$. Then 
\begin{align*}
    \widehat{\textrm{cov}}(\widehat{\Bar{\beta}}_{\scriptscriptstyle\mathrm{ML\_I}})={\frac{1}{n}}\left\{\widehat{I}^*_{11}{+}\Sigma_{\lambda}(\widehat{\Bar{\beta}}_{\scriptscriptstyle\mathrm{ML\_I}})\right\}^{-\text{\scriptsize 1}}\widehat{I}^*_{11}\left\{\widehat{I}^*_{11}{+}\Sigma_{\lambda}(\widehat{\Bar{\beta}}_{\scriptscriptstyle\mathrm{ML\_I}})\right\}^{-\textrm{\tiny T}}
\end{align*}
is used to estimate $\textrm{cov}(\widehat{\Bar{\beta}}_{\scriptscriptstyle\mathrm{ML\_I}})$ in Theorem \ref{thm2} under parametric setting (\ref{mis_pro}), {where $A^{-\textrm{\tiny T}}$ represents $(A^{-\textrm{\tiny 1}})^{\textrm{\tiny T}}$ for matrix $A$}. Let $\nabla \widehat{S}^*(\widehat{\Bar{\beta}}_{\scriptscriptstyle\mathrm{sp\_I}})$ and $\widehat{\textrm{cov}}\{\widehat{S}^*(\widehat{\Bar{\beta}}_{\scriptscriptstyle\mathrm{sp\_I}})\}$ denote the sample version of $E\{\partial \widehat{S}^*(\Bar{\beta}_{\scriptscriptstyle\mathrm{I}})/ \partial \Bar{\beta}_{\scriptscriptstyle\mathrm{I}}\}$ and $E\{\textrm{cov}\{\widehat{S}^*(\widehat{\Bar{\beta}}_{\scriptscriptstyle\mathrm{I}})\}\}$, respectively, and let $\widehat{\Sigma}_{\scriptscriptstyle\mathrm{sp}}$ represent the sample version of $\Sigma_{\scriptscriptstyle\mathrm{sp}}$. Then, {under semiparametric setting (\ref{ga_10_01}),} $\textrm{cov}(\widehat{\Bar{\beta}}_{\scriptscriptstyle\mathrm{sp\_I}})$ in Theorem \ref{thm5} is estimated by 
\begin{align*}
    \widehat{\textrm{cov}}(\widehat{\Bar{\beta}}_{\scriptscriptstyle\mathrm{sp\_I}})=&\left\{\nabla \widehat{S}^*(\widehat{\Bar{\beta}}_{\scriptscriptstyle\mathrm{sp\_I}})-n\Sigma_{\lambda}(\widehat{\Bar{\beta}}_{\scriptscriptstyle\mathrm{sp\_I}})\right\}^{-\textrm{\scriptsize 1}}\Big[\widehat{\textrm{cov}}\{S^*(\widehat{\Bar{\beta}}_{\scriptscriptstyle\mathrm{sp\_I}})\}+{n}\frac{(1-\delta)^2}{\delta^2}\widehat{\Sigma}_{\scriptscriptstyle\mathrm{sp}}\Big]\\
    &\cdot\left\{\nabla \widehat{S}^*(\widehat{\Bar{\beta}}_{\scriptscriptstyle\mathrm{sp\_I}})-n\Sigma_{\lambda}(\widehat{\Bar{\beta}}_{\scriptscriptstyle\mathrm{sp\_I}})\right\}^{-\textrm{\tiny T}}.\notag
\end{align*}

\section{Implementation procedures}\label{sec.4}

\subsection{Optimization algorithm}\label{sec.4.1}

Maximizing (\ref{Q*_ml}) to find estimator $\widehat{\theta}_{\scriptscriptstyle\mathrm{ML}}$ or finding estimator $\widehat{\Bar{\beta}}_{\scriptscriptstyle\mathrm{sp}}$ in (\ref{beta_sp_es}) is basically an optimization problem that can be written as
\begin{align}\label{opt_pbm}
    \min_{\beta} \left\{\mathcal{F}(\beta)+\mathcal{P}_{\lambda}(\beta)\right\},
\end{align}
where $\mathcal{F}(\beta)$ is a loss function taken as negative log-likelihood (\ref{log_lik_para}) or negative semiparametric log-likelihood (\ref{semi_log_lik}), and $\mathcal{P}_{\lambda}(\beta)=\sum_{j=1}^{p}{\rho_{\lambda}}(|\beta_j|)$ is a penalty function with regularization parameter $\lambda$. Due to the involvement of the measurement error probabilities, the loss function $\mathcal{F}(\beta)$ is not necessarily convex. We write $\mathcal{P}_{\lambda}(\beta)$ as the sum of the $L_1$ penalty and a nonconvex part, denoted $\mathcal{Q}_{\lambda}(\beta)$, that is, $\mathcal{P}_{\lambda}(\beta) = \lambda \|\beta\|_1 + \mathcal{Q}_{\lambda}(\beta)$, where {$\|a\|_1\triangleq\sum_{j=1}^{r}|a_j|$} represents the $L_1$ norm of vector {$a=(a_1,...,a_r)^{\textrm{\tiny T}}$}. Let $\tilde{\mathcal{F}}(\beta)$ denote $\mathcal{F}(\beta)+\mathcal{Q}_{\lambda}(\beta)$. Then the optimization problem (\ref{opt_pbm}) can be reformulated as 
\begin{align}\label{opt_rfm}
    \min_{\beta} \left\{\tilde{\mathcal{F}}_{\lambda}(\beta)+\lambda \|\beta\|_1\right\},
\end{align}
which is a nonconvex regularized optimization problem with the $L_1$ penalty.

To solve (\ref{opt_rfm}), we use the approximate regularization path-following {(APF)} method  proposed by \cite{wang2014optimal}. Consider a decreasing sequence of regularization parameters $\{\lambda_t\}_{t=0}^{N}$, where $\lambda_0$ is the smallest regularization parameter corresponding to an all-zero solution and $\lambda_N$ is the target regularization parameter, which is proportional to $\sqrt{\frac{\log p}{n}}$. For each $\lambda_t$, a variant of Nesterov's proximal-gradient method \cite{nesterov2013gradient} is iteratively used to obtain an approximate local solution until a pre-specified stopping criterion is satisfied. In the $k$th iteration of the proximal method, we minimize the local quadratic approximation given by
\begin{align}
    \Psi_{L_t^k,\lambda_t}(\beta;\beta_t^{k-1})=\tilde{\mathcal{F}}_{\lambda_t}(\beta_t^{k-1})+{\big\{}\nabla\tilde{\mathcal{F}}_{\lambda_t}(\beta_t^{k-1}){\big\}^{\mathrm{\scriptscriptstyle T}}}(\beta-\beta_t^{k-1})+\frac{L_t^k}{2}\|\beta-\beta_t^{k-1}\|_2^2+\lambda_t\|\beta\|_1,\notag
\end{align}
where $\beta_t^{k-1}$ is the solution at iteration $(k-1)$, $L_t^k$ is a stepsize chosen by a line-search method \cite[Chapter 3]{bach2012optimization}, $\|a\|_2$ represents the $L_2$ norm of vector {$a=(a_1,\ldots,a_r)^{\text{\tiny T}}$}, and $\nabla f(\alpha)$ stands for the gradient of function $f(\alpha)$, {expressed as a column vector with the $j$th element being the partial derivative of $f(\alpha)$ with respect to the $j$th element of $\alpha$, for $j=1,\ldots,r$.}

\subsection{Tuning parameters selection}\label{sec.4.2}

The implementation of the proposed methods requires the specification of suitable tuning parameters, denoted $\Upsilon$, which {includes} the regularization parameter $\lambda$ under the parametric setting (\ref{mis_pro}) and $(\lambda, h, \omega)^{\textrm{\tiny T}}$ under the semiparametric setting (\ref{ga_10_01}). Here we consider {the use of} the generalized cross validation (GCV) statistic or the Bayesian information criterion (BIC) {for their selection, as in \cite{fan2001variable, ma2010variable, wang2007tuning}.} 

Let $\tilde{Z}=(1, \tilde{Z}_{\scriptscriptstyle\mathrm{I}}^{\textrm{\tiny T}})^{\textrm{\tiny T}}$ with $\tilde{Z}_{\scriptscriptstyle\mathrm{I}}$ denoting the {covariate vector} included in the model, and let {$\widehat{\mu}_{\scriptscriptstyle\Upsilon}(\tilde{z})=P(Y=1|\tilde{Z}=\tilde{z})$ and $\widehat{\mu}^*_{\scriptscriptstyle\Upsilon}(\tilde{z})=P(Y^*=1|\tilde{Z}=\tilde{z})$. Define ${\mathdutchcal{q}_{i}}=1/\left[{\widehat{\mu}_{\scriptscriptstyle\Upsilon}(\tilde{z_i})}{\{1-\widehat{\mu}_{\scriptscriptstyle\Upsilon}(\tilde{z_i})\}}\right]$ for $i\in\mathcal{V}$ and ${\mathdutchcal{q}_{i}}=\left\{1-\widehat{\gamma}_{01}(z_i)-\widehat{\gamma}_{10}(z_i)\right\}^2/\left[{\widehat{\mu}^*_{\scriptscriptstyle\Upsilon}(\tilde{z_i})}{\{1-\widehat{\mu}^*_{\scriptscriptstyle\Upsilon}(\tilde{z_i})\}}\right]$ for $i\in\mathcal{M\backslash V}$.} Let $\widehat{I}_{11}$ {denote} the $({ s}+1)\times({ s}+1)$ submatrix of $I^*_{\delta}(\varpi_{0})$ defined in (\ref{I_delta}) corresponding to $\bar{\beta}_{\scriptscriptstyle\mathrm{I0}}$, with $\bar{\beta}_{\scriptscriptstyle\mathrm{0}}$, {$\gamma_{01}(\cdot)$, and $\gamma_{10}(\cdot)$} replaced by { their respective estimates.} According to the expressions of {$I^*_{1\Bar{\beta}\Bar{\beta}}(\theta_0)$ and $I^*_{2\Bar{\beta}\Bar{\beta}}(\theta_0)$ in Result {\color{blue}1} in the Supplementary Material, we estimate $\widehat{I}_{11}$ by $\frac{1}{n}\sum_{i=1}^{n}\mathdutchcal{q}_{i}\left\{\frac{\partial {\widehat{\mu}_{\scriptscriptstyle\Upsilon}(\tilde{z_i})}}{\partial \bar{\beta}_{\scriptscriptstyle\mathrm{I}}}\right\}^{\otimes 2}$.} 

Define the effective number of parameters in the penalized equations {as}
\begin{align}\label{eq.df}
    df_{\scriptscriptstyle\Upsilon}=\textrm{trace}\left\{\widehat{I}_{11}\{\widehat{I}_{11}+\Sigma_{\lambda}(\widehat{\bar{\beta}_{\scriptscriptstyle\mathrm{I}}})\}^{\textrm{\tiny -1}}\right\},
\end{align}
{and the} deviance of the fitted model $\widehat{\mu}_{\scriptscriptstyle\Upsilon}$ as 
\begin{align*}
    D(\widehat{\mu}_{\scriptscriptstyle\Upsilon})=&2\sum_{i\in\mathcal{V}}\big({y_i}\log\{{y_i}/{\widehat{\mu}_{\scriptscriptstyle\Upsilon}(\tilde{z_i})}\}+(1-{y_i})\log[(1-{y_i})/{\{1-\widehat{\mu}_{\scriptscriptstyle\Upsilon}(\tilde{z_i})\}}]\big)\\
    &+2\sum_{i\in\mathcal{M\backslash V}}\big({y^*_i}\log\{{y^*_i}/{\widehat{\mu}^*_{\scriptscriptstyle\Upsilon}(\tilde{z_i})}\}+(1-{y^*_i})\log[(1-{y^*_i})/{\{1-\widehat{\mu}^*_{\scriptscriptstyle\Upsilon}(\tilde{z_i})\}}]\big).
\end{align*}
The generalized cross validation (GCV) statistic {and} the Bayesian information criterion (BIC) {are then defined as:}  
\begin{align*}
    GCV(\Upsilon)=\frac{D(\widehat{\mu}_{\scriptscriptstyle\Upsilon})}{n(1-df_{\scriptscriptstyle\Upsilon}/n)^2}\ \ \text{and}\ \ BIC(\Upsilon)=D(\widehat{\mu}_{\scriptscriptstyle\Upsilon})+2(\log n) df_{\scriptscriptstyle\Upsilon},
\end{align*}
respectively. Let $\Gamma$ denote {a} grid of values for $\Upsilon$. Then the GCV criterion is to find $\Upsilon_{GCV}=\arg\min_{\Upsilon\in\Gamma}GCV(\Upsilon)$, and the BIC criterion finds $\Upsilon_{BIC}=\arg\min_{\Upsilon\in\Gamma}BIC(\Upsilon)$, {which can} be used {to implement our} the proposed methods.

\subsection{Implementation details}\label{sec.4.3}

\subsubsection{Parametric method}\label{sec.4.3.1}

{For the parametric method considered in Section \ref{sec.3}}, the parameter of interest {$\Bar{\beta}=(\beta^*_0,\beta^{\text{\tiny T}})^{\text{\tiny T}}$} and the nuisance parameter $\nu$ are iteratively optimized until the changes in the updated estimates of {$\Bar{\beta}$} and $\nu$ are small enough; in our experiment, we use $10^{-6}$ as {a} threshold {to assess the differences between two consecutive updated estimates of the parameters}, which usually takes about 5 iterations {in our numerical studies}. To be specific, we initiate $\beta$ to be 0. In each iteration, we first fix $\beta$ as the optimization value of the previous iteration, and use the {Python} package \texttt{SciPy} to implement the optimization procedures for {$\beta^*_0$ and} $\nu$. Then, we fix {$\beta^*_0$ and} $\nu$ at this value and estimate $\beta$ by employing the approximate regularization path-following method described in Section \ref{sec.4.1}.

To determine an optimal value for the tuning parameter $\lambda$, set $\Gamma=\{\lambda_1,...,\lambda_N\}$ with $\lambda_t=\varsigma^{t}\lambda_0$ for $t=0,...,N$, where $N$ is a positive integer specified by the user, $\varsigma$ is usually taken as a constant from $[0.9,1)$, and $\lambda_0$ is an initial regularization parameter chosen as $\lambda_0 = \big\|-\frac{\partial}{\partial\beta}\ell^*(\theta)\big|_{\beta=0,\ {\beta^*_0=\beta^*_{0,\text{current}},\ \nu=\nu_{\text{current}}}}^{\scriptscriptstyle\mathrm{T}}\big\|_{\infty}$ \cite[Proposition 1.2]{wang2014optimal,bach2012optimization}, with {the nuisance parameter $\nu$ and the intercept $\beta^*_0$ in $\ell^*(\theta)$} taken as the values {$\nu_{\text{current}}$ and $\beta^*_{0,\text{current}}$} in the current iteration. Here $\ell^*(\theta)$ is the log-likelihood function given in (\ref{log_lik_para}), and $\|a\|_{\infty}$ is the infinity norm of vector $a=(a_1,...,a_r)^{\textrm{\tiny T}}$ defined as $\|a\|_{\infty}=\max_{i\in\{1,...,r\}}|a_i|$. To determine a feasible value of $N$, we use the discussion of \cite{wang2014optimal} who took the target regularization parameter to be proportional to $\sqrt{\log {p}/n}$ and set $\lambda_N=c\sqrt{\log {p}/n}$ for a constant $c$, which yields $N=\lceil\frac{\log(c\sqrt{\log {p}/n}/\lambda_0)}{\log\varsigma}\rceil$. Here we take $c=\tfrac{1}{2}$. Then an optimal value of $\lambda$ is chosen as the one that minimizes the GCV or BIC over $\Gamma$ as described in Section \ref{sec.4.2}. 

\subsubsection{Semiparametric method}\label{sec.4.3.2}

To implement the strategy under the semiparametric setting (\ref{ga_10_01}), we take the kernel function in (\ref{K_b}) to be the Gaussian kernel given by ${K_{h}(u)}=\frac{1}{\sqrt{2\pi}{h}}{\exp\left(-\frac{u^2}{2{h}^2}\right)}$ with bandwidth $h$ \cite{hardle1990applied}. Following \cite{racine2004nonparametric}, we set $h$ to be proportional to {$n_v^{-1/(4+{p_1})}$}, the smoothing parameter $w$ to be proportional to {$n_v^{-2/(4+{p_1})}$}. To be specific, we take 10 values at equal distances between {$0.5n_v^{-1/(4+{p_1})}$ and $2n_v^{-1/(4+{p_1})}$} as candidate values for $h$, and take 5 values equally spaced between {$0.5n_v^{-2/(4+{p_1})}$ and $2n_v^{-2/(4+{p_1})}$} as candidate values for $w$. The sequence of regularization parameters, $\{\lambda_0,...,\lambda_N\}$, is selected in the same way as described above for the parametric setting, except that the objective function $\ell^*(\theta)$ is replaced by the semiparametric log-likelihood function $\widehat{\ell}^*({\Bar{\beta}})$ defined in (\ref{semi_log_lik}), i.e., the initial regularization parameter is taken as $\lambda_0 = \big\|-\frac{\partial}{\partial\beta}\widehat{\ell}^*({\Bar{\beta}})\big|_{\beta=0,\ {\beta^*_0=\beta^*_{0,\text{current}}}}\big\|_{\infty}$. The candidate values for $\lambda$, {$h$,} and $w$ described previously form the grid set $\Gamma$, and the optimal tuning parameters $\Upsilon=(\lambda, h, \omega)^{\textrm{\tiny T}}$ are chosen by minimizing GCV or BIC over $\Gamma$, as described in Section \ref{sec.4.2}.

{As commented in Appendix {\color{blue}B} in the Supplementary Material, nonparametric density estimators converge to the true density {slowly} as ${p_1}$ increases, a phenomenon commonly referred to as the curse of dimensionality. To mitigate this issue, we may introduce an additional step by  first applying principal component analysis (PCA) to perform dimensionality reduction on the continuous covariates \cite{eckstein2023dimensionality}. The kernel estimator in (\ref{ga_10_01}) is then applied to the transformed data. This method is evaluated in our simulation studies.} 

\section{Numerical studies}\label{sec.5}

\subsection{Simulation study}\label{sec.5.1}

\subsubsection{Simulation design}\label{sec.5.1.1}

We now conduct simulation studies to evaluate the performance of the proposed methods in contrast to the naive method. For each parameter configuration, we repeat simulation $M\triangleq 200$ times by using a method described below to analyze simulated data. We set the sample size for the primary sample $\{\{Z_i, Y^*_i\}:i\in\mathcal{M}\}$ to be $n=1000$ and take the size for the validation subsample $\{\{Z_i, Y^*_i, Y_i\}:i\in\mathcal{V}\subset\mathcal{M}\}$ as $n_v=\lceil \delta n\rceil$ with $\delta=0.1$, $0.3$ or $0.5$. 

First, we generate covariates {$Z_i=(Z_{i1},..,{Z_{i,p-1},Z_{ip}})^{\scriptscriptstyle\mathrm{T}}$} independently for $i=1,...,n$. Specifically, $(Z_{i1},...{Z_{i,p-2}})^{\scriptscriptstyle\mathrm{T}}$ is generated from the normal distribution with mean $0$ and {a} covariance {structure with $\text{Cov}(Z_{ij},Z_{ik})=0.5^{|j-l|}$}, and {$Z_{i,p-1}$ and $Z_{ip}$ are} independently generated from the Bernoulli distribution, Bernoulli(0.5). Next, given $Z_i$, {we specify model (\ref{Model}) as a logistic regression model and} generate the response $Y_i$ from model (\ref{Model}) independently for $i=1,...,n$, where {we set ${p}=20$, $\beta_0=(2, 1.3, 0, 0, 2, -1.5, 0, 0, 0, 1, 0, ..., 0)^{\scriptscriptstyle\mathrm{T}}$, and $\beta^*_{00}=1$}. Finally, conditioned on ${Z_i=z_i}$ and ${Y_i=y_i}$, surrogate measurements $Y^*_i$ are generated independently for $i=1,...,n$ from the Bernoulli distribution, Bernoulli($\gamma({z_i})$), where $\gamma({z_i})=\left\{1-\gamma_{10}({z_i})\right\}{y_i}+\gamma_{01}({z_i})\cdot(1-{y_i})$, with 
\begin{align}\label{mis_noise}
    \gamma_{01}({z_i})=\gamma_{10}({z_i})=\eta F(Z_{i2}^2)+(1-\eta)\frac{\exp (\alpha_0+{\alpha_1^{\mathrm{\scriptscriptstyle T}}{z_i}})}{1+\exp (\alpha_0+{\alpha_1^{\mathrm{\scriptscriptstyle T}}{z_i}})}.
\end{align}
{Here,} $F(\cdot)$ is the cumulative distribution function of the normal distribution $\mathcal{N}(\varrho,1)$, with $\varrho$ denoting the mean parameter, and  {$\eta\in[0,1]$} is a constant {controlling the degree of model misspecification.} Here, $\eta=0$ corresponds to {a} correctly specified {model}, {and} $\eta\ne 0$ indicates {model} misspecification { in the parametric method}. 

{For the generated data, we evaluate the performance of the proposed parametric and semiparametric methods under different levels of model misspecification, as controlled by $\eta$. To ensure a fair comparison, we maintain a consistent average misclassification rate (AMR) across different levels of model misspecification, thereby isolating the effects of model misspecification without introducing confounding variations in AMR. Specifically, \textcolor{black}{for the $j$th simulation, let ${y_i}^{*\scriptscriptstyle(j)}$, ${y_i}^{\scriptscriptstyle(j)}$ and ${z_i}^{\scriptscriptstyle(j)}=({z}_{i1}^{\scriptscriptstyle(j)}, ..., {z}_{ip}^{\scriptscriptstyle(j)})^{\textrm{\tiny T}}$ denote the simulated {values} for $Y_i^{*}$, $Y_i$, and $Z_i$ {for} $i=1,...,n$ and $j=1,...,M$. We {define} the misclassification rate (MR) {for the $j$th simulation} as $MR^{\scriptscriptstyle(j)}=\frac{1}{n}\sum_{i=1}^{n}I({y}^{*\scriptscriptstyle(j)}_i\ne {y_i}^{\scriptscriptstyle(j)})$. The AMR is then computed as $AMR=\frac{1}{M}\sum_{j=1}^{M}MR^{\scriptscriptstyle(j)}$.} To ensure a consistent AMR across different levels of model misspecification, we first fix $\alpha_0$ and $\alpha_1$, which determine the AMR contribution of the logistic component in (\ref{mis_noise}). We then select $\varrho$ such that $\mathcal{N}(\varrho,1)$ in (\ref{mis_noise}) contributes approximately an equivalent proportion to the AMR. Based on this setup, we consider the following five types of misclassification probabilities:}
\begin{align}
    &\textrm{(I)}\ \eta=0,\ \alpha_0={-2.15},\ \alpha_1={(1, 1, -1.5, 1.1, -1.3, 0, \ldots, 0)^{\mathrm{\scriptscriptstyle T}}};\notag\\
    &\textrm{(II)}\ \eta=0.5,\ \alpha_0={-2.15},\ \alpha_1={(1, 1, -1.5, 1.1, -1.3, 0, \ldots, 0)^{\mathrm{\scriptscriptstyle T}}},\ \varrho=1.98;\notag\\
    &\textrm{(III)}\ \eta=1.0,\ \alpha_0={-2.15},\ \alpha_1={(1, 1, -1.5, 1.1, -1.3, 0, \ldots, 0)^{\mathrm{\scriptscriptstyle T}}},\ \varrho=1.98;\notag\\
    &\textrm{(IV)}\ \eta=0,\ \alpha_0={-1.01},\ \alpha_1={(1, 1, -1.5, 1.1, -1.3, 0, \ldots, 0)^{\mathrm{\scriptscriptstyle T}}};\notag\\
    &\textrm{(V)}\ \eta=0.5,\ \alpha_0={-1.01},\ \alpha_1={(1, 1, -1.5, 1.1, -1.3, 0, \ldots, 0)^{\mathrm{\scriptscriptstyle T}}},\ \varrho=1.33.\notag
\end{align}
The Python implementation details for {data simulation are provided} in {Appendix} {\color{blue}E1} of the Supplementary Material. {The AMR} for settings (I), (II), and (III), is {approximately $22\%$, while for} settings (IV) and (V), {it} is around {$36\%$}. 

{Furthermore, we conduct additional simulation studies by varying the sample size $n$ and the link function in model (\ref{Model}), generating the five types of mismeasured responses described above. The results of these additional simulation studies are presented in {Appendix} {\color{blue}E} of the Supplementary Material.
} 

\subsubsection{Analysis methods}\label{sec.6.1.2}

We analyze the simulated data using the following three strategies. The first strategy evaluates the effects of ignoring the misclassification feature by {applying} the penalized logistic method {directly} to the observed measurements $\{\{Z_i, Y^*_i\}:i\in\mathcal{M}\}$; the second and third strategies implement the proposed methods with parametric and semiparametric models (\ref{mis_pro}) and (\ref{ga_10_01}) assumed respectively, {where the semiparametric model employs kernel estimators, with or without PCA, for misclassification probability estimation.} In implementing each of these strategies, we take the {penalty function to be either SCAD or MCP, which satisfies} the regularity conditions in {Appendix} {\color{blue}B4} of the Supplementary Material; in comparison, we also employ each strategy with the Lasso {(i.e., $L_1$)} penalty. We denote the resulting { methods as ``Naive - $L_1$'', ``Naive - SCAD'', ``Naive - MCP'', ``Parametric - $L_1$'', ``Parametric - SCAD'', ``Parametric - MCP'', ``Semiparametric (Kernel) - $L_1$'', ``Semiparametric (pcaKernel) - $L_1$'', ``Semiparametric (Kernel) - SCAD'', ``Semiparametric (pcaKernel) - SCAD'', ``Semiparametric (Kernel) - MCP'', and ``Semiparametric (pcaKernel) - MCP'', respectively, where ``Naive - $L_1$'',  ``Naive - SCAD'' and ``Naive - MCP''} are implemented with the {Python} package \texttt{pycasso} \cite{ge2019picasso} for the optimization procedures, and the optimization method described in Section \ref{sec.4.1} is used for estimation of $\beta$ for the proposed methods with GCV or BIC used to select tuning parameters. 

When implementing the {second} strategy under the parametric model, we take (\ref{mis_pro}) to be of the form {
\begin{align}\label{eq.exp_mis}
    \gamma_{01}(z_i)=\frac{\exp(\nu_{1}+\nu_{2}^{\textrm{\scriptsize T}} z_i)}{1+\exp(\nu_{1}+\nu_{2}^{\textrm{\scriptsize T}} z_i)}\ \text{and}\ \gamma_{10}(z_i)=\frac{\exp(\nu_{3}+\nu_{4}^{\textrm{\scriptsize T}} z_i)}{1+\exp(\nu_{3}+\nu_{4}^{\textrm{\scriptsize T}} z_i)}
\end{align}
to fit the data, where $\nu=(\nu_1,\nu_2^{\textrm{\tiny T}}, \nu_3, \nu_4^{\textrm{\tiny T}})^{\textrm{\tiny T}}$} is a vector of parameters. Thus, settings (I) and (IV) in Section \ref{sec.5.1.1} with $\eta=0$ reflect that the misclassification models are correctly specified, while settings (II), (III) and (V) present cases with misspecified classification models to different degrees.

\subsubsection{Performance metrics}\label{sec.6.1.3}

{To evaluate the overall performance of the proposed methods against the naive methods, we consider the prediction error}, defined as the average error of the prediction of $Y$, given $Z$ (e.g., \cite{fan2001variable}): $PE(\widehat{\mu})=E\left[\{Y-\widehat{\mu}(Z)\}^2\right]$, where $\widehat{\mu}(Z)$ represents the predicted value for the response based on the covariate $Z$, and the expectation is taken with respect to the joint distribution of $Y$ and $Z$. Alternatively, the prediction error can be written as $PE(\widehat{\mu})=E\left[\{Y-\mu(Z)\}^2\right]+E\left[\{\mu(Z)-\widehat{\mu}(Z)\}^2\right]$, with the first component showing the error caused by the random noise of the data and second component, called the \textit{model error} and denoted $ME(\widehat{\mu})$, reflecting the variability induced from using a model or method. Thus, the model error $ME(\widehat{\mu})$ can be used to assess the performance of a model. {For instance,} with the logistic regression model, we have that $\mu({z})=\exp(\beta^*_0+{z}^{\scriptscriptstyle\mathrm{T}}\beta)/\{1+\exp(\beta^*_0+{z}^{\scriptscriptstyle\mathrm{T}}\beta)\}$ and $\widehat{\mu}({z})=\exp(\widehat{\beta}^*_0+{z}^{\scriptscriptstyle\mathrm{T}}\widehat{\beta})/\{1+\exp(\widehat{\beta}^*_0+{z}^{\scriptscriptstyle\mathrm{T}}\widehat{\beta})\}$, where {$\widehat{\beta}^*_0$ and} $\widehat{\beta}$ represent the estimator of model parameter{s $\beta^*_{0}$ and $\beta$ obtained} from an estimation method.

{Specifically,} using the same model in Section \ref{sec.5.1.1}, we generate a new set of covariates of a large size,  ${\cal T}=\left\{z_i:i\in\mathcal{S}\right\}$ with $|\mathcal{S}|=10^4$. For simulation $j=1,..., M$, let $\widehat{\Bar{\beta}}^{(j)}$ denote the estimate of ${\Bar{\beta}}$ and let {$\widehat{\mu}^{(j)}({z_i})$} denote the resulting estimate of $\mu({z_i})$ determined by model (\ref{Model}) for a given {$Z_i=z_i$}; and calculate an approximate model error defined {above}, given by $\widehat{ME}^{(j)}\triangleq\frac{1}{|\mathcal{S}|}\sum_{i\in\mathcal{S}}\{{\mu(z_i)}-{\widehat{\mu}^{(j)}(z_i)}\}^2$, where ${\mu(z_i)}$ is given by model (\ref{Model}) with the true parameter value for {$\Bar{\beta}$}. Then we calculate the average of $\{\widehat{ME}^{(j)}: j=1,...,M\}$, and called it an ``approximate model error'' (AME). 

Next, we consider the average number of correctly recognized zero coefficients (denoted ``True-Zero'') and that of  correctly recognized nonzero coefficients (denoted ``True-NonZero'') among those $M$ simulations. These measures have been commonly used in the literature (e.g., \cite{fan2001variable, ma2010variable, liang2009variable}). To see how these values differ from the truth, we report the difference between True-Zero and {$p-s=15$} (denoted ``False-Zero'') and the difference between True-NonZero and {$s=5$} (Denoted ``False-NonZero''), {where $s$ denotes the dimension of the nonzero parameters among the ${p}$-dimensional regression coefficients, as defined in Section \ref{sec.2}}. 

To assess the performance of parameter estimation, we report the estimation results for each of nonzero coefficients, say $\beta_k$, by summarizing the results derived from $M$ simulations. In particular, let $\widehat{\beta}_k^{(j)}$ denote the estimate of $\beta_k$ from the $j$th simulation. We calculate the average of the difference (denoted ``bias'') between the estimate $\widehat{\beta}_k^{(j)}$ and the true value $\beta_{k0}$ for $j=1,...,M$, {given by} $bias(\beta_k)\triangleq\frac{1}{M}\sum_{j=1}^{M}(\widehat{\beta}_k^{(j)}-\beta_{k0})$; the sample standard deviation (denoted ``ESD'') of $\{\widehat{\beta}_k^{(j)}:j=1,...,M\}$, i.e., {$\sqrt{\frac{1}{M-1}\sum_{j=1}^{M}(\widehat{\beta}_k^{(j)}-\Bar{\beta}_k)^2}$,} with $\Bar{\beta}_k=\frac{1}{M}\sum_{j=1}^{M}\widehat{\beta}_k^{(j)}$; and the mean squared error (denoted ``MSE''), given by $\frac{1}{M}\sum_{j=1}^{M}(\widehat{\beta}_k^{(j)}-\beta_{k0})^2$. Further, let $SE_k^{(j)}$ denote the square root of the asymptotic variance for the estimator $\widehat{\beta}_k^{(j)}$ that is obtained from applying Theorems \ref{thm2} and \ref{thm5} to the $j$th simulated dataset, respectively, corresponding to the proposed parametric and semiparametric methods, and calculate the coverage rate (``CR'') of 95\% confidence intervals for $\beta_k$ that capture $\beta_k$, i.e., $C_k/M$, where $C_k$ is the number of 95\% confidence intervals $\left\{(\widehat{\beta}_k^{(j)}-1.96SE_k^{(j)}, \widehat{\beta}_k^{(j)}+1.96SE_k^{(j)}):j=1,...,M\right\}$ that contain {$\beta_{k0}$}.

\subsubsection{Simulation results}\label{sec.simulation_result}\label{sec.6.1.4}

To save space, here we report the simulation results for settings {(I)-(II)} only and defer the results for {settings (III), (IV), and (V)} in {Appendix} {\color{blue}E} of the Supplementary Material. Table \ref{table.20_1000_AME} presents the results for the \textit{AME} (multiplied by $10^2$), False-NonZero and False-Zero. Clearly, the naive method yields noticeably  biased results. On the contrary, the proposed parametric and semiparametric methods produce substantially improved results with significantly smaller AME values than those of the naive method as well as close-to-zero values of False-NonZero and False-Zero. While the two proposed methods outperform the naive method, the performance of these two methods differ. The proposed parametric method has better performance than the proposed semiparametric method when the model is correctly specified (i.e., $\eta=0$), whereas the proposed semiparametric method outperforms the proposed parametric method in the presence of model misspecification (i.e., $\eta\ne 0$). Comparing the use of the SCAD, { MCP} or $L_1$ penalty, the biased performance of the naive method does not appear to profoundly differ; however, the performance of the two proposed methods does depend on the choice of the penalty function. The use of the SCAD {and MCP} evidently yields better results than the use of the $L_1$ penalty, and this agrees with the theoretical results in Section \ref{sec.3} for which the SCAD {and MCP} satisfy the required regularity conditions but the $L_1$ penalty does not. 

In the absence of model misspecification, the proposed parametric method outperforms the proposed semiparametric method, reflecting by yielding smaller bias and ESD values than the proposed semiparametric method. On the contrary, with model misspecification, the proposed semiparametric method offers us protection by outputting better results than the proposed parametric method. {However, as the validation ratio increases, the negative impact of model misspecification on the parametric method diminishes.} Tables \ref{table.20_1000_bias_I} and \ref{table.20_1000_bias_II} report on the estimation results for those nonzero coefficients (i.e., {$\beta_1$, $\beta_2$, $\beta_5$, $\beta_6$, and $\beta_{10}$}) obtained from using the {GCV} tuning parameter; the results obtained from the {BIC} criterion {are} reported in {Appendix} {\color{blue}E} of the Supplementary Material. Evidently, the naive method {incurs} considerably biases, whereas the proposed parametric and semiparametric estimators always produce remarkably smaller finite sample biases. While such improvement is at the cost of inflating the variability of the estimators (i.e., ESD values for the proposed methods are larger than those for the naive method), the overall benefits of the proposed methods are evident by the smaller values of MSE than those for the naive method. Comparing the use of the SCAD, {MCP} and $L_1$ penalty functions, it is clear again that the estimators utilizing the $L_1$ penalty are severely biased, even if the proposed methods are used and the validation ratio is high, which is in accordance with the theoretical results in Section \ref{sec.3} since the $L_1$ penalty {{does not}} satisfy the conditions in Theorems \ref{thm1} and \ref{thm2} simultaneously as noted in Remark \ref{remark3}. 

Further, we examine the performance of the proposed models by reporting the results of CR for settings with correct model specification, where SCAD penalty {or MCP} is used. {As shown in Table \ref{table.logit_20_1000_CI},} with a good amount of validation data (e.g., $\delta=0.3$ or larger), the proposed parametric and the semiparametric methods output reasonable coverage rates that are close to the nominal value 95\%.

\newpage


{\scriptsize 
\renewcommand\arraystretch{0.3}
\setlength{\tabcolsep}{3.45mm}
\begin{longtable}{clcccccc}
\caption{Simulation results of the model error ($\times$ 100) and the number of zero coefficients.} \label{table.20_1000_AME}
\\
\toprule 
\multirow{2}*{\shortstack{Validation\\ Ratio}} &\multirow{2}*{\ \ \ \ \ \ \ \ \ Method} & \multicolumn{2}{c}{AME} & \multicolumn{2}{c}{False-NonZero} & \multicolumn{2}{c}{False-Zero}\\
\cmidrule(lr){3-4}\cmidrule(lr){5-6}\cmidrule(lr){7-8} & & GCV & BIC & GCV & BIC & GCV & BIC \\
\midrule  
\endfirsthead 
\multicolumn{8}{r}{Continued}\\ 
\toprule  
\multirow{2}*{\shortstack{Validation\\ Ratio}} &\multirow{2}*{\ \ \ \ \ \ \ \ \ Method} & \multicolumn{2}{c}{AME} & \multicolumn{2}{c}{False-NonZero} & \multicolumn{2}{c}{False-Zero}\\
\cmidrule(lr){3-4}\cmidrule(lr){5-6}\cmidrule(lr){7-8} & & GCV & BIC & GCV & BIC & GCV & BIC \\
\midrule  
\endhead 
\bottomrule    
\multicolumn{8}{c}{Next Page}\\  
\endfoot 
\bottomrule 
\endlastfoot
\multicolumn{8}{c}{{Setting} (I) with $\eta=0$ and AMR $\approx$ 0.22} \\
\cmidrule(lr){1-8}
& Naive - SCAD & \textbf{6.109} & \textbf{7.224} & \textbf{3.900} & 
\textbf{0.655} & \textbf{0.140} & \textbf{0.855} \\
\cmidrule(lr){2-8}
$\delta=0.1$ & Parametric - SCAD & 0.958 & 3.904 & 1.915 & 0.070 & 0.130 & 1.660 \\
& Semiparametric (Kernel) - SCAD & 1.612 & 3.340 & 1.545 & 0.100 & 0.275 & 1.310 \\
& Semiparametric (pcaKernel) - SCAD & 1.737 & 3.324 & 1.320 & 0.120 & 0.345 & 1.300 \\
\cmidrule(lr){2-8}
$\delta=0.3$ & Parametric - SCAD & 0.467 & 0.852 & 2.190 & 0.155 & 0.000 & 0.295 \\
& Semiparametric (Kernel) - SCAD & 0.621 & 1.246 & 1.865 & 0.185 & 0.040 & 0.490 \\
& Semiparametric (pcaKernel) - SCAD & 0.603 & 1.639 & 1.655 & 0.145 & 0.045 & 0.705 \\
\cmidrule(lr){2-8}
$\delta=0.5$ & Parametric - SCAD & 0.330 & 0.324 & 2.065 & 0.090 & 0.005 & 0.080 \\
& Semiparametric (Kernel) - SCAD & 0.340 & 0.369 & 1.470 & 0.175 & 0.010 & 0.085 \\
& Semiparametric (pcaKernel) - SCAD & 0.369 & 0.349 & 1.575 & 0.130 & 0.015 & 0.070  \\
\specialrule{0em}{1pt}{1pt}
\cmidrule(lr){1-8}
& Naive - MCP & \textbf{6.120} & \textbf{7.224} & \textbf{3.805} & \textbf{0.615} & \textbf{0.145} & \textbf{0.915}  \\
\cmidrule(lr){2-8}
$\delta=0.1$ & Parametric - MCP & 0.894 & 4.550 & 1.795 & 0.085 & 0.130 & 1.875 \\
& Semiparametric (Kernel) - MCP & 1.707 & 3.160 & 0.050 & 1.460 & 1.220 & 0.350\\
& Semiparametric (pcaKernel) - MCP & 1.971 & 3.862 & 1.085 & 0.070 & 0.505 & 1.510 \\
\cmidrule(lr){2-8}
$\delta=0.3$ & Parametric - MCP & 0.453 & 0.992 & 2.250 & 0.160 & 0.010 & 0.385\\
& Semiparametric (Kernel) - MCP & 0.603 & 1.467 & 1.860 & 0.170 & 0.030 & 0.600\\
& Semiparametric (pcaKernel) - MCP & 0.679 & 1.819 & 1.950 & 0.115 & 0.070 & 0.805 \\
\cmidrule(lr){2-8}
$\delta=0.5$ & Parametric - MCP & 0.330 & 0.304 & 2.205 & 0.070 & 0.000 & 0.080 \\
& Semiparametric (Kernel) - MCP & 0.383 & 0.408 & 1.920 & 0.150 & 0.005 & 0.095 \\
& Semiparametric (pcaKernel) - MCP & 0.376 & 0.369 & 1.770 & 0.175 & 0.010 & 0.065 \\
\specialrule{0em}{1pt}{1pt}
\cmidrule(lr){1-8}
& Naive - $L_1$ & \textbf{7.449} & \textbf{8.525} & \textbf{4.325} & \textbf{1.855} & \textbf{0.495} & \textbf{0.755}  \\
\cmidrule(lr){2-8}
$\delta=0.1$ & Parametric - $L_1$ & 1.249 & 6.342 & 10.695 & 0.575 & 0.065 & 2.445 \\
& Semiparametric (Kernel) - $L_1$ & 3.271 & 3.869 & 3.105 & 1.405 & 0.350 & 0.740 \\
& Semiparametric (pcaKernel) - $L_1$ & 3.242 & 3.898 & 3.030 & 1.375 & 0.380 & 0.785 \\
\cmidrule(lr){2-8}
$\delta=0.3$ & Parametric - $L_1$ & 0.666 & 2.081 & 10.880 & 3.280 & 0.005 & 0.445\\
& Semiparametric (Kernel) - $L_1$ & 2.630 & 2.970 & 2.785 & 1.555 & 0.170 & 0.305\\
& Semiparametric (pcaKernel) - $L_1$ & 2.628 & 3.018 & 2.880 & 1.530 & 0.175 & 0.355 \\
\cmidrule(lr){2-8}
$\delta=0.5$ & Parametric - $L_1$ & 0.878 & 0.954 & 11.91 & 3.900 & 0.000 & 0.055 \\
& Semiparametric (Kernel) - $L_1$ & 1.605 & 1.844 & 2.060 & 1.280 & 0.055 & 0.070 \\
& Semiparametric (pcaKernel) - $L_1$ & 1.621 & 1.891 & 1.945 & 1.305 & 0.040 & 0.095\\
\specialrule{0em}{1pt}{1pt}
\cmidrule(lr){1-8}
\multicolumn{8}{c}{{Setting} (II) with  $\eta=0.5$ and AMR $\approx$ 0.22} \\
\cmidrule(lr){1-8}
& Naive - SCAD & \textbf{7.336} & \textbf{7.870} & \textbf{3.190} & \textbf{0.180} & \textbf{0.415} & \textbf{1.020} \\
\cmidrule(lr){2-8}
$\delta=0.1$ & Parametric - SCAD & 3.060 & 8.331 & 1.510 & 0.040 & 0.755 & 3.120\\
& Semiparametric (Kernel) - SCAD & 2.102 & 3.805 & 1.285 & 0.045 & 0.550 & 1.570\\
& Semiparametric (pcaKernel) - SCAD & 2.124 & 4.394 & 1.095 & 0.090 & 0.610 & 1.700 \\
\cmidrule(lr){2-8}
$\delta=0.3$ & Parametric - SCAD & 0.623 & 1.191 & 2.220 & 0.200 & 0.010 & 0.415 \\
& Semiparametric (Kernel) - SCAD & 0.500 & 1.254 & 1.725 & 0.120 & 0.015 & 0.550 \\
& Semiparametric (pcaKernel) - SCAD & 0.551 & 1.330 & 1.785 & 0.100 & 0.040 & 0.595 \\
\cmidrule(lr){2-8}
$\delta=0.5$ & Parametric - SCAD & 0.383 & 0.358 & 2.135 & 0.140 & 0.000 & 0.080 \\
& Semiparametric (Kernel) - SCAD & 0.349 & 0.327 & 1.600 & 0.145 & 0.005 & 0.065 \\
& Semiparametric (pcaKernel) - SCAD & 0.323 & 0.339 & 1.430 & 0.095 & 0.005 & 0.080 \\
\specialrule{0em}{1pt}{1pt}
\cmidrule(lr){1-8}
& Naive - MCP & \textbf{7.331} & \textbf{7.787} & \textbf{2.995} & \textbf{0.150} & \textbf{0.440} & \textbf{1.105}  \\
\cmidrule(lr){2-8}
$\delta=0.1$ & Parametric - MCP & 3.353 & 8.045 & 1.600 & 0.025 & 0.975 & 2.985\\
& Semiparametric (Kernel) - MCP & 2.134 & 4.284 & 1.120 & 0.060 & 0.615 & 1.755 \\
& Semiparametric (pcaKernel) - MCP & 2.175 & 4.179 & 0.930 & 0.085 & 0.640 & 1.710 \\
\cmidrule(lr){2-8}
$\delta=0.3$ & Parametric - MCP & 0.590 & 1.167 & 2.136 & 0.145 & 0.005 & 0.425 \\
& Semiparametric (Kernel) - MCP & 0.510 & 1.374 & 1.765 & 0.125 & 0.020 & 0.610 \\
& Semiparametric (pcaKernel) - MCP & 0.602 & 1.416 & 1.805 & 0.105 & 0.060 & 0.640 \\
\cmidrule(lr){2-8}
$\delta=0.5$ & Parametric - MCP & 0.368 & 0.370 & 2.075 & 0.165 & 0.000 & 0.085 \\
& Semiparametric (Kernel) - MCP & 0.347 & 0.309 & 1.705 & 0.085 & 0.000 & 0.070 \\
& Semiparametric (pcaKernel) - MCP & 0.360 & 0.369 & 1.545 & 0.100 & 0.010 & 0.105 \\
\specialrule{0em}{1pt}{1pt}
\cmidrule(lr){1-8}
& Naive - $L_1$ & \textbf{8.322} & \textbf{9.529} & \textbf{3.675} & \textbf{1.545} & \textbf{0.945} & \textbf{1.380}  \\
\cmidrule(lr){2-8}
$\delta=0.1$ & Parametric - $L_1$ & 3.224 & 9.053 & 8.475 & 0.395 & 0.580 & 2.875\\
& Semiparametric (Kernel) - $L_1$ & 2.965 & 3.698 & 4.180 & 1.580 & 0.235 & 0.580\\
& Semiparametric (pcaKernel) - $L_1$ & 2.970 & 3.708 & 4.165 & 1.240 & 0.205 & 0.585 \\
\cmidrule(lr){2-8}
$\delta=0.3$ & Parametric - $L_1$ & 1.142 & 2.833 & 10.470 & 2.580 & 0.020 & 0.520\\
& Semiparametric (Kernel) - $L_1$ & 2.031 & 2.377 & 3.280 & 1.650 & 0.080 & 0.165 \\
& Semiparametric (pcaKernel) - $L_1$ & 2.129 & 2.251 & 3.060 & 1.865 & 0.090 & 0.135 \\
\cmidrule(lr){2-8}
$\delta=0.5$ & Parametric - $L_1$ & 0.798 & 0.797 & 11.335 & 3.640 & 0.000 & 0.040 \\
& Semiparametric (Kernel) - $L_1$ & 1.338 & 1.654 & 2.425 & 1.520 & 0.005 & 0.045 \\
& Semiparametric (pcaKernel) - $L_1$ & 1.374 & 1.666 & 2.295 & 1.640 & 0.015 & 0.040 \\
\end{longtable}
}


\begin{landscape}

{\scriptsize
\renewcommand\arraystretch{0.3}
\begingroup
\setlength{\LTleft}{-20cm plus -1fill}
\setlength{\LTright}{\LTleft}
\begin{longtable}{clcccccccccc}
\caption{Simulation results of bias (ESD) and MSE of the estimators for nonzero coefficients: {Setting} (I) with $\eta=0$ and AMR $\approx$ 0.22.} \label{table.20_1000_bias_I}
\\
\toprule  
\multirow{2}*{\shortstack{Validation\\ Ratio}} & \multirow{2}*{\ \ \ \ \ \ \ Method}  & \multicolumn{2}{c}{$\beta_1$} & \multicolumn{2}{c}{$\beta_2$} & \multicolumn{2}{c}{$\beta_5$} & \multicolumn{2}{c}{$\beta_6$} & \multicolumn{2}{c}{$\beta_{10}$}\\
\cmidrule(lr){3-4}\cmidrule(lr){5-6}\cmidrule(lr){7-8}\cmidrule(lr){9-10}\cmidrule(lr){11-12} & & bias (ESD) & MSE & bias (ESD) & MSE & bias (ESD) & MSE & bias (ESD) & MSE & bias (ESD) & MSE \\
\midrule 
\endfirsthead 
\multicolumn{12}{r}{Continued}\\ 
\toprule 
\multirow{2}*{\shortstack{Validation\\ Ratio}} & \multirow{2}*{\ \ \ \ \ \ \ Method}  & \multicolumn{2}{c}{$\beta_1$} & \multicolumn{2}{c}{$\beta_2$} & \multicolumn{2}{c}{$\beta_5$} & \multicolumn{2}{c}{$\beta_6$} & \multicolumn{2}{c}{$\beta_{10}$}\\
\cmidrule(lr){3-4}\cmidrule(lr){5-6}\cmidrule(lr){7-8}\cmidrule(lr){9-10}\cmidrule(lr){11-12} & & bias (ESD) & MSE & bias (ESD) & MSE & bias (ESD) & MSE & bias (ESD) & MSE & bias (ESD) & MSE \\
\midrule 
\endhead
\bottomrule    
\multicolumn{12}{c}{Next Page}\\  
\endfoot 
\bottomrule 
\endlastfoot 
& Naive - SCAD & \textbf{-1.626\ (0.082)} & \textbf{2.652} & \textbf{-1.125\ (0.103)} & \textbf{1.275} & \textbf{-1.262\ (0.095)} & \textbf{1.602} & \textbf{ 1.103\ (0.084)} & \textbf{1.223} & \textbf{-0.748\ (0.093)} & \textbf{0.567} \\
\cmidrule(lr){2-12}
$\delta=0.1$ & Parametric - SCAD  & \ 0.174 (0.444) & 0.226 & \ 0.162 (0.335) &  0.138 & \ 0.133 (0.462) & 0.230 & -0.082 (0.411) & 0.175 & \ 0.041 (0.358) & 0.129 \\
\cmidrule(lr){3-12} 
& Semiparametric (Kernel) - SCAD & -0.077 (0.530) & 0.286 & -0.045 (0.446) & 0.200 & -0.048 (0.592) & 0.351 & \ 0.136 (0.571) & 0.342 & -0.137 (0.471) & 0.239  \\
\cmidrule(lr){3-12} 
& Semiparametric (pcaKernel) - SCAD   & -0.101 (0.528) & 0.287 & -0.107 (0.497) & 0.258 & -0.079 (0.637) & 0.410 & \ 0.170 (0.606) & 0.395 & -0.142 (0.456) & 0.227  \\
\cmidrule(lr){2-12}
$\delta=0.3$ & Parametric - SCAD & \ 0.169 (0.281) & 0.107 &
\ 0.148 (0.243) & 0.081 &
\ 0.134 (0.252) & 0.081 &
-0.069 (0.250) & 0.067 &
\ 0.072 (0.222) & 0.054 \\
\cmidrule(lr){3-12} 
& Semiparametric (Kernel) - SCAD  & \ 0.218 (0.311) & 0.144 & \ 0.149 (0.270) & 0.095 & \ 0.208 (0.340) & 0.158 & -0.101 (0.344) & 0.128 & \ 0.090 (0.275) & 0.084 \\
\cmidrule(lr){3-12} 
& Semiparametric (pcaKernel) - SCAD  & \ 0.206 (0.315) & 0.141 & \ 0.138 (0.268) & 0.091 & \ 0.192 (0.336) & 0.149 & \ 0.092 (0.320) & 0.110 & \ 0.074 (0.284) & 0.086  \\
\cmidrule(lr){2-12}
$\delta=0.5$ & Parametric - SCAD  & \ 0.077 (0.218) & 0.053 &
\ 0.069 (0.194) & 0.042 &
\ 0.056 (0.204) & 0.044 &
-0.031 (0.203) & 0.042 &
\ 0.036 (0.190) & 0.037 \\
\cmidrule(lr){3-12} 
& Semiparametric (Kernel) - SCAD  & \ 0.104 (0.230) & 0.063 & \ 0.083 (0.196) & 0.045 & \ 0.102 (0.255) & 0.075 & -0.063 (0.216) & 0.050 & \ 0.047 (0.206) & 0.044  \\
\cmidrule(lr){3-12} 
& Semiparametric (pcaKernel) - SCAD  & \ 0.107 (0.244) & 0.071 & \ 0.077 (0.228) & 0.058 & \ 0.094 (0.251) & 0.072 & -0.049 (0.214) & 0.048 & \ 0.038 (0.212) & 0.046 \\
\specialrule{0em}{3pt}{3pt}
\cmidrule(lr){1-12}
& Naive - MCP  & \textbf{-1.626 (0.082)} & \textbf{2.651} & \textbf{-1.126 (0.104)} & \textbf{1.279} & \textbf{-1.263 (0.096)} & \textbf{1.604} & \textbf{ 1.104 (0.084)} & \textbf{1.225} & \textbf{-0.748 (0.093)} & \textbf{0.568} \\
\cmidrule(lr){2-12}
$\delta=0.1$ & Parametric - MCP  & \ 0.189 (0.360) & 0.164 & \ 0.153 (0.351) & 0.146 & \ 0.106 (0.499) & 0.259 & -0.084 (0.435) & 0.195 & \ 0.022 (0.362) & 0.131 \\
\cmidrule(lr){3-12} 
& Semiparametric (Kernel) - MCP   & -0.077 (0.558) & 0.316 & -0.072 (0.482) & 0.237 & -0.056 (0.626) & 0.393 & \ 0.133 (0.595) & 0.370 & -0.136 (0.475) & 0.243  \\
\cmidrule(lr){3-12} 
& Semiparametric (pcaKernel) - MCP  & -0.145 (0.561) & 0.334 & -0.146 (0.531) & 0.302 & -0.156 (0.626) & 0.415 & \ 0.243 (0.655) & 0.486 & -0.221 (0.501) & 0.298  \\
\cmidrule(lr){2-12}
$\delta=0.3$ & Semiparametric (Kernel) - MCP  & \ 0.126 (0.381) & 0.160 &
\ 0.088 (0.370) & 0.144 &
-0.045 (0.536) & 0.288 &
\ 0.106 (0.547) & 0.309 &
-0.054 (0.401) & 0.163 \\
\cmidrule(lr){3-12} 
& Semiparametric (pcaKernel) - MCP  & \ 0.220 (0.283) & 0.128 & \ 0.157 (0.262) & 0.093 & \ 0.219 (0.337) & 0.161 & -0.099 (0.347) & 0.130 & \ 0.099 (0.254) & 0.074  \\
\cmidrule(lr){3-12} 
& Semiparametric (pcaKernel) - MCP  & \ 0.190 (0.299) & 0.125 & \ 0.126 (0.289) & 0.099 & \ 0.179 (0.361) & 0.161 & -0.069 (0.377) & 0.146 & \ 0.063 (0.302) & 0.095 \\
\cmidrule(lr){2-12}
$\delta=0.5$ & Semiparametric (Kernel) - MCP  & \ 0.078 (0.221) & 0.055 &
\ 0.068 (0.193) & 0.042 &
\ 0.057 (0.214) & 0.049 &
-0.029 (0.209) & 0.044 &
\ 0.035 (0.181) & 0.034 \\
\cmidrule(lr){3-12} 
& Semiparametric (pcaKernel) - MCP  & \ 0.107 (0.237) & 0.067 & \ 0.085 (0.210) & 0.051 & \ 0.099 (0.246) & 0.070 & -0.055 (0.210) & 0.047 & \ 0.054 (0.202) & 0.044 \\
\cmidrule(lr){3-12} 
& Semiparametric (pcaKernel) - MCP & \ 0.102 (0.240) & 0.068 & \ 0.078 (0.214) & 0.051 & \ 0.094 (0.251) & 0.071 & -0.053 (0.214) & 0.048 & \ 0.039 (0.209) & 0.045  \\

\specialrule{0em}{3pt}{3pt}
\cmidrule(lr){1-12}
&Naive - $L_1$  & \textbf{-1.691 (0.075)} & \textbf{2.867} & \textbf{-1.148 (0.078)} & \textbf{1.323} & \textbf{-1.503 (0.099)} & \textbf{2.268} & \textbf{ 1.360 (0.141)} & \textbf{1.87} & \textbf{-0.834 (0.072)} & \textbf{0.701}  \\
\cmidrule(lr){2-12}
$\delta=0.1$ & Parametric - $L_1$  & \ 0.128 (0.492) & 0.258 & \ 0.141 (0.419) & 0.194 & -0.134 (0.470) &  0.237 & \ 0.142 (0.402) & 0.181 & -0.082 (0.307) &  0.101 \\
\cmidrule(lr){3-12} 
& Semiparametric (Kernel) - $L_1$  & -0.528 (0.543) & 0.572 & -0.308 (0.416) & 0.267 & -0.777 (0.485) & 0.837 & \ 1.048 (0.357) & 1.225 & -0.612 (0.313) & 0.472  \\
\cmidrule(lr){3-12} 
& Semiparametric (pcaKernel) - $L_1$  & -0.561 (0.516) & 0.579 & -0.344 (0.408) & 0.284  & -0.773 (0.456) & 0.803 & \ 1.057 (0.353) & 1.240 & -0.618 (0.307) & 0.475  \\
\cmidrule(lr){2-12}
$\delta=0.3$ & Parametric - $L_1$  & -0.005 (0.297) & 0.088 &
\ 0.024 (0.276) & 0.076 &
-0.128 (0.258) & 0.083 &
\ 0.154 (0.265) & 0.094 &
-0.069 (0.214) & 0.05  \\
\cmidrule(lr){3-12} 
& Semiparametric (Kernel) - $L_1$    & \ 0.283 (0.443) & 0.275 & \ 0.309 (0.382) & 0.241 & -0.604 (0.360) & 0.493 & \ 0.935 (0.330) & 0.982 & -0.529 (0.307) & 0.374  \\
\cmidrule(lr){3-12} 
& Semiparametric (pcaKernel) - $L_1$  & \ 0.288 (0.414) & 0.254 & \ 0.310 (0.367) & 0.230 & -0.589 (0.346) & 0.467 & \ 0.945 (0.318) & 0.993 & -0.532 (0.309) & 0.379  \\
\cmidrule(lr){2-12}
$\delta=0.5$ & Parametric - $L_1$  & 
\ 0.182 (0.358) & 0.160 &
\ 0.164 (0.325) & 0.132 &
\ 0.056 (0.302) & 0.094 &
\ 0.106 (0.248) & 0.073 &
\ 0.011 (0.210) & 0.044 \\
\cmidrule(lr){3-12} 
& Semiparametric (Kernel) - $L_1$  & -0.208 (0.389) & 0.194 & -0.137 (0.296) & 0.106 & -0.898 (0.225) & 0.856 & \ 0.826 (0.239) & 0.739 & -0.448 (0.192) & 0.237 \\
\cmidrule(lr){3-12} 
& Semiparametric (pcaKernel) - $L_1$  & -0.224 (0.406) & 0.214 & -0.154 (0.313) & 0.121 & -0.905 (0.227) & 0.871 & \ 0.810 (0.233) & 0.710 & -0.465 (0.196) & 0.255  \\
\end{longtable}
\endgroup
}


{\scriptsize 
\renewcommand\arraystretch{0.3}
\begingroup
\setlength{\LTleft}{-20cm plus -1fill}
\setlength{\LTright}{\LTleft}
\begin{longtable}{clcccccccccc}
\caption{Simulation results of bias (ESD) and MSE of the estimators for nonzero coefficients: {Setting} (II) with $\eta=0.5$ and AMR $\approx$ 0.22.}\label{table.20_1000_bias_II} \\
\toprule  
\multirow{2}*{\shortstack{Validation\\ Ratio}} & \multirow{2}*{\ \ \ \ \ \ \ Method}  & \multicolumn{2}{c}{$\beta_1$} & \multicolumn{2}{c}{$\beta_2$} & \multicolumn{2}{c}{$\beta_5$} & \multicolumn{2}{c}{$\beta_6$} & \multicolumn{2}{c}{$\beta_{10}$}\\
\cmidrule(lr){3-4}\cmidrule(lr){5-6}\cmidrule(lr){7-8}\cmidrule(lr){9-10}\cmidrule(lr){11-12} & & bias (ESD) & MSE & bias (ESD) & MSE & bias (ESD) & MSE & bias (ESD) & MSE & bias (ESD) & MSE \\
\midrule 
\endfirsthead 
\multicolumn{12}{r}{Continued}\\ 
\toprule  
\multirow{2}*{\shortstack{Validation\\ Ratio}} & \multirow{2}*{\ \ \ \ \ \ \ Method}  & \multicolumn{2}{c}{$\beta_1$} & \multicolumn{2}{c}{$\beta_2$} & \multicolumn{2}{c}{$\beta_5$} & \multicolumn{2}{c}{$\beta_6$} & \multicolumn{2}{c}{$\beta_{10}$}\\
\cmidrule(lr){3-4}\cmidrule(lr){5-6}\cmidrule(lr){7-8}\cmidrule(lr){9-10}\cmidrule(lr){11-12} & & bias (ESD) & MSE & bias (ESD) & MSE & bias (ESD) & MSE & bias (ESD) & MSE & bias (ESD) & MSE \\
\midrule 
\endhead
\bottomrule    
\multicolumn{12}{c}{Next Page}\\  
\endfoot 
\bottomrule 
\endlastfoot 
& {{Naive - SCAD}} &  \textbf{-1.544 (0.091)} & \textbf{2.392} & \textbf{-1.396 (0.110)} & \textbf{1.959} & \textbf{-1.355 (0.082)} & \textbf{1.842} & \textbf{ 1.086 (0.083)} & \textbf{1.187} & \textbf{-0.732 (0.086)} & \textbf{0.544} \\
\cmidrule(lr){2-12}
$\delta=0.1$ & {Parametric - SCAD} & 
-0.177 (0.723) & 0.551 &
\ 0.304 (0.858) & 0.824 &
-0.352 (0.843) & 0.831 &
\ 0.305 (0.662) & 0.529 &
\ -0.235 (0.481) & 0.285\\
\cmidrule(lr){3-12} 
& Semiparametric (Kernel) - SCAD  & -0.136 (0.569) & 0.341 & -0.516 (0.640) & 0.674 & -0.101 (0.652) & 0.433 & \ 0.174 (0.581) & 0.366 & -0.157 (0.428) & 0.207 \\
\cmidrule(lr){3-12} 
& Semiparametric (pcaKernel) - SCAD &  -0.102 (0.578) & 0.343 & -0.553 (0.668) & 0.750 & -0.053 (0.620) & 0.386 & \ 0.145 (0.573) & 0.348 & -0.157 (0.460) & 0.235 \\
\cmidrule(lr){2-12}
$\delta=0.3$ & {{Parametric - SCAD}} & 
\ 0.159 (0.284) & 0.106 &
\ 0.423 (0.281) & 0.257 & 
\ 0.123 (0.281) & 0.094 &
-0.054 (0.265) & 0.073 &
\ 0.053 (0.231) & 0.056\\
\cmidrule(lr){3-12} 
& Semiparametric (Kernel) - SCAD & \ 0.216 (0.287) & 0.129 & \ 0.046 (0.285) & 0.083 & \ 0.222 (0.299) & 0.138 & -0.127 (0.269) & 0.088 & \ 0.092 (0.258) & 0.075 \\
\cmidrule(lr){3-12} 
& Semiparametric (pcaKernel) - SCAD & \ 0.216 (0.281) & 0.125 & \ 0.024 (0.292) & 0.086 & \ 0.206 (0.310) & 0.138 & -0.116 (0.303) & 0.105 & \ 0.078 (0.265) & 0.076 \\
\cmidrule(lr){2-12}
$\delta=0.5$ & {{Parametric - SCAD}} & 
\ 0.093 (0.219) & 0.056 &
\ 0.227 (0.195) & 0.090 &
\ 0.059 (0.225) & 0.054 &
-0.040 (0.199) & 0.041 &
\ 0.046 (0.199) & 0.041\\
\cmidrule(lr){3-12} 
& Semiparametric (Kernel) - SCAD &  \ 0.127 (0.248) & 0.077 & \ 0.035 (0.222) & 0.050 & \ 0.101 (0.244) & 0.069 & -0.058 (0.208) & 0.047 & \ 0.042 (0.197) & 0.040  \\
\cmidrule(lr){3-12} 
& Semiparametric (pcaKernel) - SCAD &  \ 0.121 (0.242) & 0.073 & \ 0.027 (0.216) & 0.047 & \ 0.097 (0.252) & 0.072 & -0.058 (0.213) & 0.049 & \ 0.038 (0.199) & 0.041 \\
\specialrule{0em}{3pt}{3pt}
\cmidrule(lr){1-12}
& {{Naive - MCP}} & \textbf{-1.545 (0.090)} & \textbf{2.396} & \textbf{-1.393 (0.110)} & \textbf{1.953} & \textbf{-1.355 (0.083)} & \textbf{1.842} & \textbf{ 1.087 (0.084)} & \textbf{1.189} & \textbf{-0.733 (0.087)} & \textbf{0.545} \\
\cmidrule(lr){2-12}
$\delta=0.1$ & {Parametric - MCP} & 
-0.206 (0.704) & 0.535 &
\ 0.280 (0.823) & 0.753 &
-0.462 (0.919) & 1.053 &
\ 0.405 (0.734) & 0.700 &
-0.278 (0.518) & 0.344\\
\cmidrule(lr){3-12} 
& Semiparametric (Kernel) - MCP & -0.126 (0.581) & 0.352 & -0.521 (0.671) & 0.719 & -0.082 (0.636) & 0.410 & \ 0.189 (0.600) & 0.394 & -0.173 (0.451) & 0.232 \\
\cmidrule(lr){3-12} 
& Semiparametric (pcaKernel) - MCP &  -0.130 (0.580) & 0.352 & -0.564 (0.687) & 0.788 & -0.086 (0.615) & 0.384 & \ 0.172 (0.562) & 0.343 & -0.170 (0.431) & 0.214  \\
\cmidrule(lr){2-12}
$\delta=0.3$ & {{Parametric - MCP}} & 
\ 0.160 (0.260) & 0.093 &
\ 0.425 (0.288) & 0.263 &
\ 0.116 (0.266) & 0.084 &
-0.062 (0.244) & 0.063 &
\ 0.054 (0.235) & 0.058\\
\cmidrule(lr){3-12} 
& Semiparametric (Kernel) - MCP   &  \ 0.221 (0.289) & 0.132 & \ 0.045 (0.284) & 0.082 & \ 0.224 (0.306) & 0.144 & -0.123 (0.271) & 0.088 & \ 0.087 (0.256) & 0.073 \\
\cmidrule(lr){3-12} 
& Semiparametric (pcaKernel) - MCP &  \ 0.217 (0.297) &  0.135 & \ 0.015 (0.325) & 0.106 & \ 0.207 (0.330) & 0.151 & -0.108 (0.333) & 0.122 & \ 0.070 (0.278) & 0.082 \\
\cmidrule(lr){2-12}
$\delta=0.5$ & {{Parametric - MCP}} & 
\ 0.097 (0.224) & 0.059 &
\ 0.227 (0.202) & 0.092 &
\ 0.063 (0.236) & 0.060 &
-0.040 (0.209) & 0.045 &
\ 0.042 (0.194) & 0.039\\
\cmidrule(lr){3-12} 
& Semiparametric (Kernel) - MCP  &  \ 0.135 (0.237) & 0.074 & \ 0.043 (0.223) & 0.051 & \ 0.117 (0.247) & 0.074 & -0.068 (0.211) & 0.049 & \ 0.057 (0.176) & 0.034 \\
\cmidrule(lr){3-12} 
& Semiparametric (pcaKernel) - MCP &  \ 0.132 (0.238) & 0.074 & \ 0.033 (0.214) & 0.047 & \ 0.107 (0.245) & 0.071 & -0.057 (0.234) & 0.058 & \ 0.049 (0.191) & 0.039  \\

\specialrule{0em}{3pt}{3pt}
\cmidrule(lr){1-12}
&{{Naive - $L_1$}} &  \textbf{-1.656 (0.075)} & \textbf{2.749} & \textbf{-1.321 (0.041)} & \textbf{1.747} & \textbf{-1.557 (0.081)} & \textbf{2.432} & \textbf{ 1.313 (0.148)} & \textbf{1.746} & \textbf{-0.821 (0.072)} & \textbf{0.679}  \\
\cmidrule(lr){2-12}
$\delta=0.1$ & {{Parametric - $L_1$}} & 
-0.300 (0.701) & 0.579 &
\ 0.303 (0.765) & 0.675 &
-0.573 (0.742) & 0.875 &
\ 0.561 (0.509) & 0.572 &
-0.309 (0.405) & 0.259\\
\cmidrule(lr){3-12} 
& Semiparametric (Kernel) - $L_1$ &  -0.619 (0.471) & 0.603 & -0.889 (0.362) & 0.921 & -0.707 (0.456) & 0.707 & \ 0.851 (0.344) & 0.842 & -0.532 (0.263) & 0.351 \\
\cmidrule(lr){3-12} 
& Semiparametric (pcaKernel) - $L_1$ &  -0.567 (0.511) & 0.582 & -0.872 (0.372) & 0.899 & -0.683 (0.489) & 0.705 & \ 0.857 (0.332) & 0.844 & -0.507 (0.279) & 0.334 \\
\cmidrule(lr){2-12}
$\delta=0.3$ & {{Parametric - $L_1$}} & 
\ 0.072 (0.387) & 0.154 &
\ 0.343 (0.353) & 0.241 &
-0.180 (0.288) & 0.115 &
\ 0.25 (0.313) & 0.160 &
-0.107 (0.231) & 0.064\\
\cmidrule(lr){3-12} 
& Semiparametric (Kernel) - $L_1$ & \ 0.162 (0.579) & 0.359 & -0.102 (0.465) & 0.225 & -0.607 (0.373) & 0.506 & \ 0.820 (0.318) & 0.773 & -0.455 (0.272) & 0.281 \\
\cmidrule(lr){3-12} 
& Semiparametric (pcaKernel) - $L_1$ &  \ 0.164 (0.575) & 0.356 & -0.117 (0.458) & 0.223 & -0.596 (0.380) & 0.499 & \ 0.841 (0.315) & 0.805 & -0.479 (0.278) & 0.306  \\
\cmidrule(lr){2-12}
$\delta=0.5$ & {{Parametric - $L_1$}} & 
\ 0.103 (0.324) & 0.115 &
\ 0.237 (0.306) & 0.149 &
-0.001 (0.312) & 0.097 &
\ 0.132 (0.227) & 0.069 &
-0.022 (0.202) & 0.041\\
\cmidrule(lr){3-12} 
& Semiparametric (Kernel) - $L_1$ &  -0.301 (0.414) & 0.261 & -0.316 (0.277) & 0.176 & -0.844 (0.240) & 0.769 & \ 0.770 (0.224) & 0.643 & -0.443 (0.170) & 0.225 \\
\cmidrule(lr){3-12} 
& Semiparametric (pcaKernel) - $L_1$ &  -0.291 (0.436) & 0.274 & -0.321 (0.284) & 0.183 & -0.849 (0.226) & 0.772 & \ 0.779 (0.236) & 0.662 & -0.451 (0.182) & 0.236 \\
\end{longtable}
\endgroup
}

\end{landscape}


{\scriptsize 
\renewcommand\arraystretch{0.3}
\setlength{\tabcolsep}{12pt} 
\begingroup
\begin{longtable}{clccccc}
\caption{Simulation results for the proposed methods with SCAD or MCP using the GCV tuning parameter selector: coverage rates of 95\% confidence intervals for nonzero coefficients.}\label{table.logit_20_1000_CI}\\    
\toprule  
Validation Ratio & \hspace{1cm}Method & $\beta_1$ &  $\beta_2$ &  $\beta_5$ & $\beta_6$ & $\beta_{10}$ \\
\midrule  
\endfirsthead 
\multicolumn{7}{r}{Continued}\\      
\toprule  
Validation Ratio & \hspace{1cm}Method & $\beta_1$ &  $\beta_2$ &  $\beta_5$ & $\beta_6$ & $\beta_{10}$ \\
\midrule 
\endhead 
\bottomrule    
\multicolumn{7}{c}{Next Page}\\  
\endfoot 
\bottomrule 
\endlastfoot  
\multicolumn{7}{c}{{Setting} (I) with AMR $\approx$ 0.22} \\
\cmidrule(lr){1-7}
$\delta=0.1$ & Parametric - SCAD & 0.909 & 0.889 & 0.960 & 0.924 & 0.848 \\
& Semiparametric (Kernel) - SCAD & 0.917 & 0.933 & 0.928 & 0.883 & 0.839 \\
& Semiparametric (pcaKernel) - SACD & 0.906 & 0.906 & 0.889 & 0.883 & 0.822 \\
\cmidrule(lr){2-7}
$\delta=0.3$ & Parametric - SCAD & 0.935 & 0.920 & 0.945 & 0.925 & 0.880 \\
& Semiparametric (Kernel) - SCAD & 0.972 & 0.967 & 0.950 & 0.950 & 0.944 \\
& Semiparametric (pcaKernel) - SACD & 0.944 & 0.983 & 0.950 & 0.961 & 0.939 \\
\cmidrule(lr){2-7}
$\delta=0.5$ & Parametric - SCAD & 0.980 & 0.945 & 0.970 & 0.940 & 0.920  \\
& Semiparametric (Kernel) - SCAD & 0.972 & 0.961 & 0.967 & 0.950 & 0.928 \\
& Semiparametric (pcaKernel) - SACD & 0.961 & 0.939 & 0.967 & 0.967 & 0.917 \\
\cmidrule(lr){1-7}
$\delta=0.1$ & Parametric - MCP & 0.925 & 0.865 & 0.905 & 0.885 & 0.845 \\
& Semiparametric (Kernel) - MCP & 0.883 & 0.922 & 0.894 & 0.883 & 0.817 \\
& Semiparametric (pcaKernel) - MCP & 0.861 & 0.872 & 0.867 & 0.800 & 0.739\\
\cmidrule(lr){2-7}
$\delta=0.3$ & Parametric - MCP & 0.935 & 0.920 & 0.950 & 0.930 & 0.885 \\
& Semiparametric (Kernel) - MCP & 0.994 & 0.978 & 0.956 & 0.944 & 0.950\\
& Semiparametric (pcaKernel) - MCP &0.983 & 0.956 & 0.933 & 0.933 & 0.922  \\
\cmidrule(lr){2-7}
$\delta=0.5$ & Parametric - MCP & 0.980 & 0.960 & 0.965 & 0.930 & 0.910 \\
& Semiparametric (Kernel) - MCP & 0.967 & 0.967 & 0.961 & 0.956 & 0.906 \\
& Semiparametric (pcaKernel) - MCP & 0.961 & 0.939 & 0.961 & 0.956 & 0.928 \\
\specialrule{0.05em}{3pt}{1pt}
\specialrule{0.05em}{0pt}{3pt}
\multicolumn{7}{c}{{Setting} (IV) with AMR $\approx$ 0.36} \\
\cmidrule(lr){1-7}
$\delta=0.1$ & Parametric - SCAD & 0.905 & 0.915 & 0.930 & 0.935 & 0.895 \\
& Semiparametric (Kernel) - SCAD & 0.900 & 0.806 & 0.761 & 0.717 & 0.594 \\
& Semiparametric (pcaKernel) - SCAD &  0.872 & 0.789 & 0.689 & 0.589 & 0.478 \\
\cmidrule(lr){2-7}
$\delta=0.3$ & Parametric - SCAD & 0.950 & 0.930 & 0.975 & 0.950 & 0.905 \\
& Semiparametric (Kernel) - SCAD & 0.994 & 0.989 & 0.967 & 0.967 & 0.944 \\
& Semiparametric (pcaKernel) - SCAD & 0.983 & 0.983 & 0.956 & 0.956 & 0.900\\
\cmidrule(lr){2-7}
$\delta=0.5$ & Parametric - SCAD & 0.965 & 0.940 & 0.960 & 0.930 & 0.940 \\
& Semiparametric (Kernel) - SCAD & 0.989 & 0.950 & 0.978 & 0.956 & 0.939 \\
& Semiparametric (pcaKernel) - SCAD & 0.983 & 0.950 & 0.956 & 0.956 & 0.917 \\
\cmidrule(lr){1-7}
$\delta=0.1$ & Parametric - MCP & 0.910 & 0.910 & 0.895 & 0.880 & 0.835 \\
& Semiparametric (Kernel) - MCP & 0.889 & 0.844 & 0.794 & 0.717 & 0.639 \\
& Semiparametric (pcaKernel) - MCP &  0.828 & 0.750 & 0.667 & 0.622 & 0.517 \\
\cmidrule(lr){2-7}
$\delta=0.3$ & Parametric - MCP & 0.945 & 0.915 & 0.960 & 0.940 & 0.925 \\
& Semiparametric (Kernel) - MCP & 0.983 & 0.978 & 0.944 & 0.944 & 0.933\\
& Semiparametric (pcaKernel) - MCP & 0.983 & 0.961 & 0.928 & 0.956 & 0.922 \\
\cmidrule(lr){2-7}
$\delta=0.5$ & Parametric - MCP & 0.985 & 0.940 & 0.965 & 0.930 & 0.935 \\
& Semiparametric (Kernel) - MCP &  0.983 & 0.950 & 0.961 & 0.944 & 0.917 \\
& Semiparametric (pcaKernel) - MCP &  0.989 & 0.950 & 0.967 & 0.950 & 0.917 \\
\end{longtable}
\endgroup
}

\subsection{A real data example}\label{sec.5.2}

{The National Health and Nutrition Examination Survey (NHANES) is a national survey that assesses the health and nutrition of adults and children in the United States. It provides publicly available data files along with related documentation (\href{https://wwwn.cdc.gov/nchs/nhanes/Default.aspx}{https://wwwn.cdc.gov/nchs/nhanes/Default.aspx}). Here, we {analyze} obesity data from the NHANES 2017–March 2020 Pre-Pandemic dataset, which includes both self-reported (mismeasured) and {clinically assessed} (true) obesity. We identify informative covariates associated with obesity, {highlighting} the consequences of ignoring response mismeasurement and demonstrating the effectiveness of our proposed methods.

The World Health Organization (WHO) defines obesity as a body mass index (BMI) greater than or equal to 30, calculated as $\text{BMI} = (\text{Weight in kg})/(\text{Height in m})^2$. We use measurements from the {\tt Body Measure} examination dataset, collected in the Mobile Examination Center (MEC) by trained health technicians, to compute BMI and determine the true obesity status for each subject. We define $Y=1$ if the corresponding BMI is greater than 30 and $Y=0$ otherwise. Additionally, we use the {\tt Weight History} questionnaire dataset, which contains self-reported height and weight, to calculate $Y^*$, the mismeasured obesity status. We consider ${p}=20$ potential predictors, denoted as $Z$, consisting of ${{p_1}}=16$ continuous and ${p_2}=4$ binary variables. These predictors capture various aspects, including body measurements, nutrient intake from dietary interviews, physical activity, and marital status. The calculations of true and mismeasured responses, along with the potential predictors, are summarized in Table {\color{blue}A77} of {Appendix} {\color{blue}E7} in the Supplementary Material. We focus on {adult} male subjects with complete predictor measurements from Table {\color{blue}A77}, resulting in a final dataset of 3237 subjects. For subject $i=1,2,\ldots,n$ with $n=3237$, let $Y_i$, $Y^*_i$, and $Z_i$ denote random variables $Y$, $Y^*$, and $Z$ corresponding to subject $i$, respectively.}

To illustrate the utility of the proposed methods and to compare {their performance}, we randomly split the data into 70\% {training} and 30\% test data, {denoting the sets of subject indices as $\mathcal{M}_{tr}$ and $\mathcal{M}_{te}$, respectively}. We use the training data to fit the models with different methods and the test data to evaluate their performance. We consider {four} methods. The first method serves as the {baseline}; it applies the usual penalized logistic regression (called ``Reference'') to the {correctly measured data} $\left\{\{Y_i, Z_i\}:i\in\mathcal{M}_{tr}\right\}$. {It is referred to as ``Reference (Pycasso)'' when solved using the {\tt Pycasso} package \cite{ge2019picasso} and ``Reference (APF)'' when using optimization method in Section \ref{sec.4.1}.} The second method, called ``Naive Method'', applies usual penalized logistic regression to surrogate response measurements and covariates, $\left\{\{Y_i^*, Z_i\}:i\in\mathcal{M}_{tr}\right\}$, which demonstrates the consequence of ignoring the differences between the surrogate and the true response. {Similar to the Reference method, we use two optimization approaches for the Naive method, referred to as ``Naive (Pycasso)'' and ``Naive (APF)'', respectively.} The {last two} methods aim to address the measurement error effects in the naive method, where the data include the main study data $\left\{\{Y_i^*, Z_i\}:i\in\mathcal{M}_{tr}\right\}$ and a small subsample of validation data $\left\{\{Y_i, Y_i^*, Z_i\}:i\in\mathcal{V}_{tr}\right\}$ with $\mathcal{V}_{tr}\subset\mathcal{M}_{tr}$ and $|\mathcal{V}_{tr}|=\lceil \delta|\mathcal{M}_{tr}|\rceil$ for different validation ratios $\delta=0.1,\ 0.3$ or 0.5. Methods {3 and 4}, respectively called ``Parametric'' and ``Semiparametric'', are the proposed methods described in Sections \ref{sec.3} and \ref{sec.3.3}. The parametric method {models misclassification probabilities $\gamma_{01}(\cdot)$ and $\gamma_{10}(\cdot)$ using logistic regression. The semiparametric method is categorized as ``Semiparametric (Kernel)'' when no PCA is performed and ``Semiparametric (pcaKernel)'' when PCA is applied. }

To assess the {method performance} from different aspects, we employ metrics that are commonly used in the error-free context and calculate them using the error-free test set $\left\{\{Y_i, Z_i\}:i\in\mathcal{M}_{te}\right\}$. {For $i\in\mathcal{M}_{te}$, we let $y_i$ and $z_i$ represent the realization of $Y_i$ and $Z_i$, respectively, and} let ${\widehat{\mu}(z_i)}$ denote the predicted probability of subject $i$ to be {obese} {given $Z_i=z_i$}. Then the predicted class {$\widehat{y}_i$ for subject $i$ is then determined} to be 1 if ${\widehat{\mu}(z_i)}>0.5$ and 0 otherwise. The accuracy (``ACC'') is the percentage of correct classifications calculated by
\begin{align*}
    ACC = \frac{1}{m}\sum_{i\in \mathcal{M}_{te}}I({y_i=\widehat{y}_i}),
\end{align*}
where $m$ denotes the number of observations in the test set. The Brier score (``Brier'') measures the squared difference between the actual response and the predicted probability, which is defined as
\begin{align*}
    Brier = \frac{1}{m}\sum_{i\in\mathcal{M}_{te}}{\{y_i-\widehat{\mu}(z_i)\}}^2.
\end{align*}
The Area Under Curve (``AUC'') indicates the ability of the model to discriminate between {obese and non-obese} samples and is estimated as
\begin{align*}
    AUC = \frac{1}{m_0 m_1}\sum_{i\in\mathcal{M}_{te}:y_i=1}\sum_{j\in\mathcal{M}_{te}:y_j=0}I{\{\widehat{\mu}(z_i)>\widehat{\mu}(z_j)\}},
\end{align*}
where $m_0$ and $m_1$ represent the number of {obese and non-obese} patients in the test set. Additionally, we report the number of selected predictors in the fitted model, denoted as ``\#Covariates''.

\begin{figure}[H]
\centering  
\subfigure[ACC]{\label{Fig.scad_gcv_logit_acc}
\includegraphics[width=7.9cm]{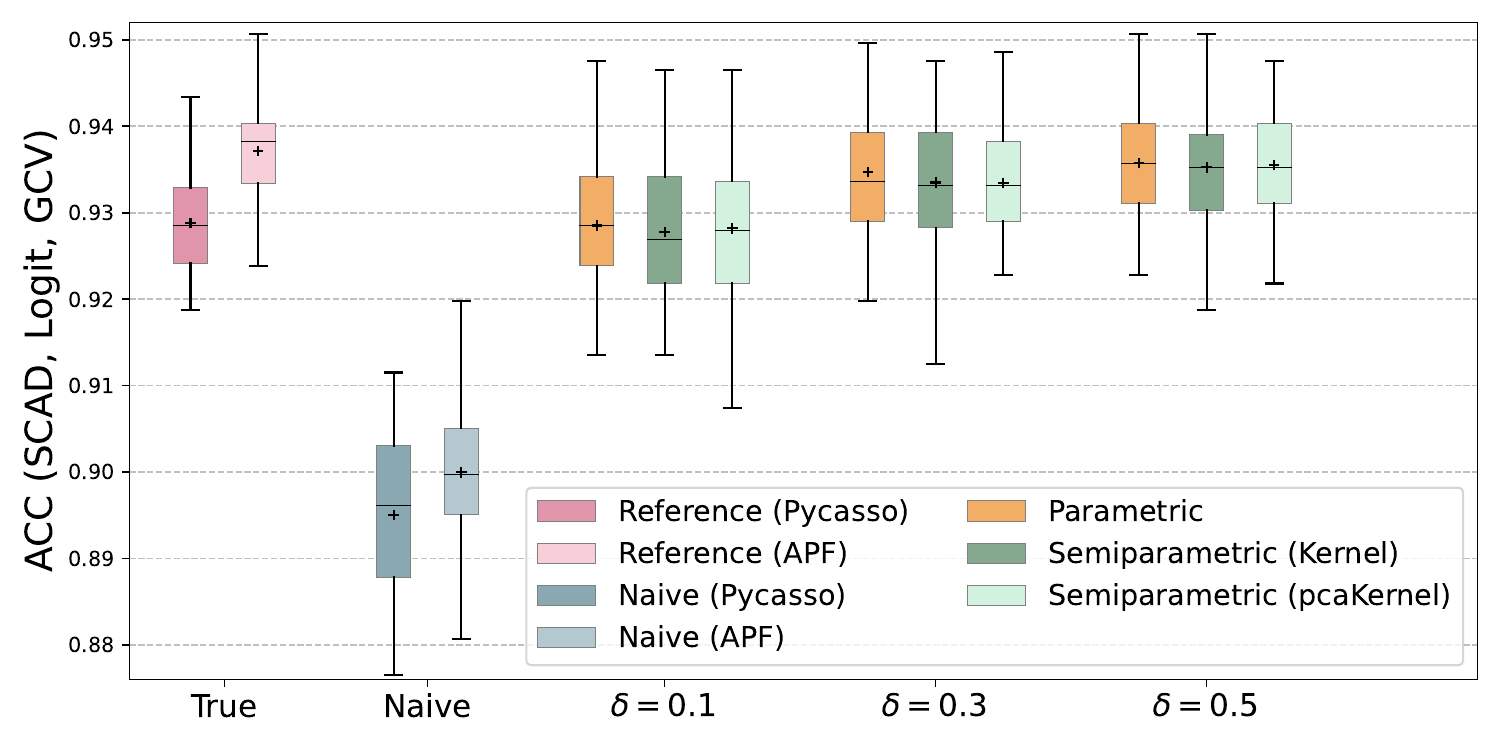}}
\subfigure[AUC]{\label{Fig.scad_gcv_logit_auc}
\includegraphics[width=7.9cm]{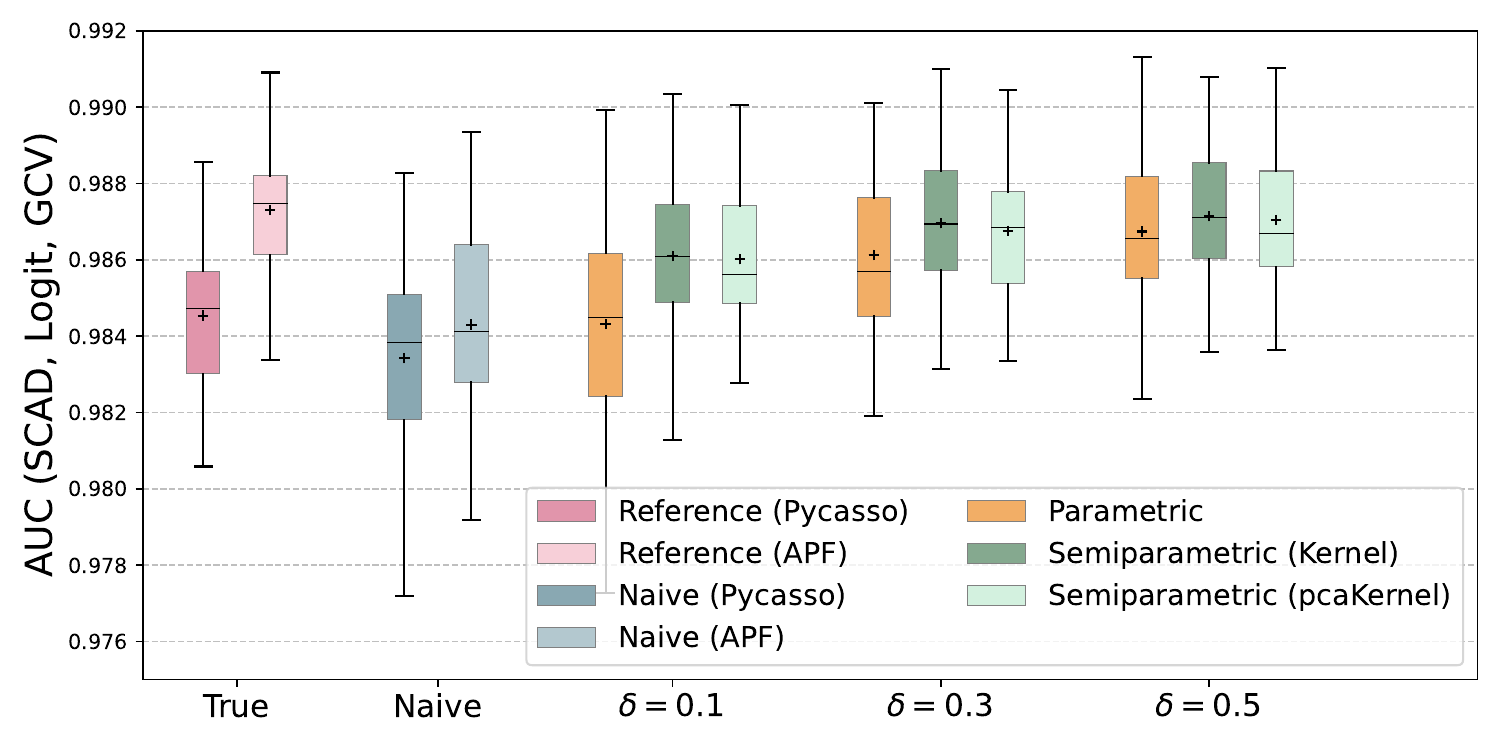}}
\subfigure[Brier]{\label{Fig.scad_gcv_logit_brier}
\includegraphics[width=7.9cm]{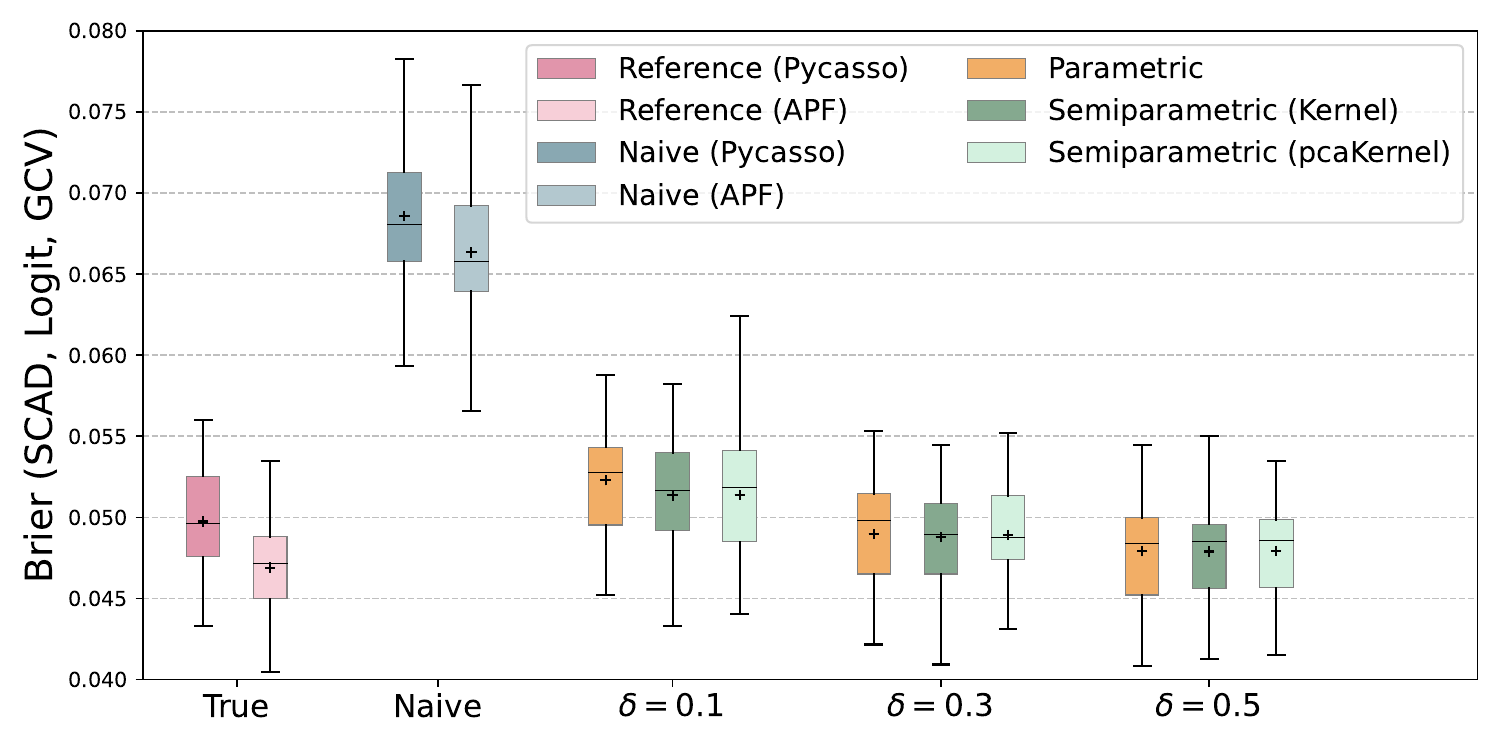}}
\caption{Boxplots of performance metrics for the NHANES obesity data, using the SCAD penalty, the GCV selector, and the logit link function for model (\ref{Model}).}
\label{Fig.scad_gcv_logit}
\end{figure}

To ameliorate the division effects of forming training and test data, we repeat the aforementioned procedure 50 times, and report in Table \ref{table.real_NHANES} the mean of each metric over 50 repetitions, as well as the corresponding standard deviations, displayed in parentheses. {In this analysis, we use a logistic regression model for model (\ref{Model}), with the SCAD penalty and the GCV tuning parameter selector. Results for other penalty and link functions are provided in {Appendix} {\color{blue}E7} in the Supplementary Material. To visualize the distribution{s} of these metric values, we show boxplots in Figure \ref{Fig.scad_gcv_logit}. It demonstrates that the unsatisfactory performance of the naive approach, significantly differing from that of the reference method. The proposed parametric and semiparametric methods noticeably improve the performance of the naive approach.}\vspace{10pt} 

\newpage

\vspace{-10pt}
{\footnotesize
\renewcommand\arraystretch{0.3}
\begin{longtable}{clcccc}      
\caption{Prediction performance of the four methods in analyzing NHANES obesity data, using the SCAD penalty, the GCV selector, and the logit link function for model (\ref{Model}).}\label{table.real_NHANES}\\  
\toprule  
Validation Ratio & \ \ \ \ \ Method & ACC & AUC & Brier & \#Covariates\\
\midrule  
\endfirsthead 
\multicolumn{6}{r}{Continued}\\      
\toprule     
Validation Ratio & \ \ \ \ \ Method & ACC & AUC & Brier & \#Covariates\\
\midrule      
\endhead 
\bottomrule    
\multicolumn{6}{c}{Next Page}\\  
\endfoot 
\bottomrule
\endlastfoot 
& Reference (Pycasso) & 0.929 (0.006) & 0.985 (0.002) & 0.050 (0.003) & \ 4.00   \\
& Reference (APF) & 0.938 (0.006) & 0.988 (0.002) & 0.044 (0.004) & \ 9.78 \\
\cmidrule(lr){1-6}
& Naive Method  (Pycasso)& 0.895 (0.009) & 0.983 (0.003) & 0.069 (0.004) & \ 4.20 \\
& Naive Method  (APF)& 0.900 (0.009) & 0.984 (0.002) & 0.066 (0.005) & 16.06 \\
\cmidrule(lr){1-6}
$\delta=0.1$ 
& Parametric & 0.911 (0.051) & 0.960 (0.137) & 0.062 (0.027) & \ 4.22 \\
& Semiparametric (Kernel) & 0.928 (0.009) & 0.986 (0.002) & 0.051 (0.004) & 13.76 \\
& Semiparametric (pcaKernel) & 0.924 (0.010) & 0.983 (0.004) & 0.054 (0.006) & \ 7.14 \\
\cmidrule(lr){1-6}
$\delta=0.3$ 
& Parametric & 0.929 (0.012) & 0.984 (0.005) & 0.050 (0.008) & \ 5.00 \\
& Semiparametric (Kernel) & 0.933 (0.008) & 0.987 (0.002) & 0.049 (0.003) & 13.18 \\
& Semiparametric (pcaKernel) & 0.929 (0.011) & 0.984 (0.004) & 0.051 (0.007) & \ 6.44\\
\cmidrule(lr){1-6}
$\delta=0.5$ 
& Parametric & 0.933 (0.010) & 0.986 (0.004) & 0.047 (0.006) & \ 6.26 \\
& Semiparametric (Kernel) & 0.935 (0.006) & 0.987 (0.002) & 0.048 (0.003) & 13.30 \\
& Semiparametric (pcaKernel) & 0.933 (0.007) & 0.986 (0.003) & 0.047 (0.005) & \ 6.52 \\
\end{longtable}
}


\section{Discussion}\label{sec.6}

Measurement error is a longstanding concern in data analysis and presents significant challenges in statistical inference. Despite available work on variable selection with error-prone covariates, relatively limited attention has been given to variable selection with error-contaminated responses. In this paper, we introduce a class of variable selection procedures for {binary classification problems} with misclassified responses. These methods utilize validation data to characterize the misclassification degree of the response variables. With properly chosen penalty functions and regularization parameters, we show that the proposed estimators are $\sqrt{n}$-consistent and have oracle properties. Furthermore, our bias correction methods are shown to be superior to naive methods that ignore the misclassification in responses.

The simulation study reveals that both parametric and semiparametric methods exhibit lower model errors and can identify more true zero coefficients than naive methods. Furthermore, in terms of estimation consistency, the proposed correction methods with the SCAD {or MCP} penalty demonstrate significant improvement, displaying notably reduced bias and mean squared errors. In particular, when the misclassification models are correctly specified, estimators generated by the parametric method achieve remarkable accuracy even with a moderate validation sample size. In cases of slight misspecification of the misclassification model, the parametric method can still perform fairly well, yet the semiparametric method exhibits greater robustness.

{While the development here is addressed to generalized linear models, it is possible to extend to other models, provided they satisfy the regularity conditions in {Appendix} {\color{blue}B1} of the Supplementary Material.} Our methods are primarily focused on binary classification, {but} they can be extended to multi-class scenarios by modifying the log-likelihood function {to incorporate generalized linear models with multi-category responses. Specifically, for a $K$-class classification problem with $K\ge 2$, where the response $Y$ and its surrogate $Y^*$ take values in $\mathcal{Y} \triangleq \{0, 1, \ldots, K-1\}$, model (\ref{Model}) can be naturally extended by defining $\mu_j(z;\Bar{\beta})\triangleq P(Y=j|Z=z)$ for $j=1,\ldots K-1$ and setting $\mu_0(z;\Bar{\beta})\triangleq 1-\sum_{j=1}^{K-1}\mu_j(z;\Bar{\beta})$, with $\Bar{\beta}$ representing the vector of all parameters involved in the response model. The misclassification probabilities in (\ref{mu_link}) can be replaced by a $K\times K$ matrix, denoted by ${\gamma}$, whose $(j+1,k+1)$ element is given by $\gamma_{j+1,k+1}(z)\triangleq P(Y^*=k|Y=j,Z=z)$ for $j,k\in\mathcal{Y}$.} Furthermore, machine learning algorithms considered by \cite{guo2023label, guo2024learning} can be utilized to estimate misclassification probabilities. Another interesting direction for future research involves exploring variable selection in cases where both covariates and responses are subject to measurement error \cite{yi2017statistical}. Additionally, it is interesting to investigate scenarios where the number of covariates {and the parameter dimension grow} to infinity as the sample size approaches infinity, {and explore} variable selection methods for high-dimensional data with measurement error \cite{datta2017cocolasso, ma2010variable}.

\section*{Acknowledgments}
Yi is the Canada Research Chair in Data Science (Tier 1). Her research was supported by the Canada Research Chairs Program and the Natural Sciences and Engineering Research Council of Canada (NSERC). {She serves as the corresponding author.}

\section*{Supplementary Material}

\noindent{\bf Supplement I}\\
Proofs of the theoretical results, {along with} additional simulation results and real data analysis. The supplement is also available at {\footnotesize\url{https://github.com/hguo1728/Variable-Selection-and-Misclassified-Responses/blob/main/Supplement%20I.pdf}}.

\noindent{\bf Supplement II}\\
Code for producing the numerical results in the manuscript. The code is also available at {\footnotesize\url{https://github.com/hguo1728/Variable-Selection-and-Misclassified-Responses}}.

\bibliographystyle{chicago}

\end{document}